\documentclass[9pt,twocolumn]{extarticle}
\pdfoutput=1
\usepackage{soul}
\usepackage{titlesec} 

\usepackage[superscript, nomove]{cite}

\usepackage{dsfont}
\usepackage{graphicx}
\usepackage{subcaption}
\usepackage[margin=0.9in]{geometry}
\usepackage[usenames,dvipsnames]{color}
\usepackage[colorlinks,linkcolor=Blue,urlcolor=Blue,citecolor=Blue]{hyperref}
\usepackage{amsmath,amssymb}
\usepackage[squaren]{SIunits}

\usepackage{mathpazo}
\usepackage{courier}
\normalfont
\usepackage[T1]{fontenc}
\usepackage{caption}
\usepackage{upgreek}

\newcommand{\ket}[1]{\ensuremath{|#1\rangle\mkern-1mu}}
\newcommand{\bra}[1]{\ensuremath{\mkern-1mu\langle#1|}}

\newcommand{\ad}[1]{\textsuperscript{#1}\kern-2pt}

\makeatletter
\def\blx@maxline{77}
\makeatother

\usepackage{capt-of}

\def\mytitle{A generalised multipath delayed-choice experiment \\on a  large-scale quantum nanophotonic chip}      

\title{\vspace{-1.0cm}\Huge\textbf{\textrm{\mytitle}}}  
 
\author{Xiaojiong Chen$^{1,\dagger}$, Yaohao Deng$^{1,\dagger}$, Shuheng Liu$^{1,\dagger}$, Tanumoy Pramanik$^{1,2}$, Jun Mao$^{1}$, Jueming Bao$^{1}$, Chonghao Zhai$^{1}$, \\Tianxiang Dai$^{1}$, Huihong Yuan$^{1}$,  Jiajie Guo$^{1}$, Shao-Ming Fei$^{3}$, Marcus Huber$^{4,5}$, Bo Tang$^{6}$, Yan Yang$^{6,\star}$, Zhihua Li$^{6}$, \\Qiongyi He$^{1,2,7,8,9,\star}$, Qihuang Gong$^{1,2,7,8,9\star}$, Jianwei Wang$^{1,2,7,8,9\star}$
}

\date{} 
\begin{document}
\twocolumn[{
\maketitle 
\vspace{-5mm}
\begin{center}
\begin{minipage}{1\textwidth}
\begin{center}
\textit{\textrm{
\textsuperscript{1} State Key Laboratory for Mesoscopic Physics, School of Physics, Peking University, Beijing, 100871, China 
\\\textsuperscript{2} Beijing Academy of Quantum Information Sciences, Beijing 100193, China
\\\textsuperscript{3} School of Mathematical Sciences, Capital Normal University, Beijing 100037, China
\\\textsuperscript{4} Institute for Quantum Optics and Quantum Information -- IQOQI Vienna, Austrian Academy of Sciences, Boltzmanngasse 3, 1090 Vienna, Austria
\\\textsuperscript{5} Vienna Center for Quantum Science and Technology, Atominstitut, TU Wien,  1020 Vienna, Austria
\\\textsuperscript{6} Institute of Microelectronics, Chinese Academy of Sciences, Beijing 100029, China
\\\textsuperscript{7} Frontiers Science Center for Nano-optoelectronics \& Collaborative Innovation Center of Quantum Matter, Peking University, Beijing, 100871, China 
\\\textsuperscript{8} Collaborative Innovation Center of Extreme Optics, Shanxi University, Taiyuan 030006, Shanxi, China
\\\textsuperscript{9} Peking University Yangtze Delta Institute of Optoelectronics, Nantong 226010, Jiangsu, China.
\\\textsuperscript{$\dagger$} These authors contributed equally to this work. \\
{$\star$} emails:  yyang10@ime.ac.cn, qiongyihe@pku.edu.cn, qhgong@pku.edu.cn, jww@pku.edu.cn\\
\vspace{5mm}
  }}
\end{center}
\end{minipage}
\end{center}

\setlength\parindent{12pt}
\begin{quotation}

{
\noindent
Famous double-slit or double-path experiments, implemented in a Young's or Mach-Zehnder interferometer, have confirmed the dual nature of quantum matter~\cite{Bohr}.  
When a stream of photons~\cite{Grangier_1986}, neutrons~\cite{RevModPhys.60.1067}, atoms~\cite{Durr1998},  or molecules~\cite{C60}, passes through two  slits, either wave-like interference fringes build up on a screen, or particle-like which-path distribution can be ascertained. 
These quantum objects  exhibit both wave and particle properties but exclusively, depending on the way they are measured~\cite{Bohr}.  
In an equivalent Mach-Zehnder configuration, the  object displays either wave or particle nature in the presence or absence of a beamsplitter, respectively, that represents the choice of which-measurement~\cite{Wootters79}. 
Wheeler further proposed a gedanken experiment~\cite{Wheeler}, in which the choice of which-measurement is delayed, \textit{i.e.} determined after the object has already entered the interferometer, so as to exclude the possibility of predicting which-measurement it will confront. The delayed-choice experiments have enabled significant demonstrations of genuine two-path duality of different quantum objects~\cite{RevModPhys.88.015005,PhysRevA.35.2532,Jacques966,PhysRevLett.100.220402}. 
Recently, a quantum controlled version of delayed-choice was proposed by Ionicioiu and Terno, by introducing a quantum-controlled beamsplitter that is in a coherent superposition of presence and absence~\cite{QBStheory}. 
It represents a controllable experiment platform that can not only reveal  wave and particle characters, but also their superposition~\cite{USTCDC,NiceDC,BristolDC,NJUDC}. 
Moreover, a quantitative description of two-slit duality relation was initialized in Wootters and Zurek's seminal work~\cite{Wootters79} and  formalized by Greenberger, Yasin, Jaeger, and Englert~\cite{Greenberger,PhysRevA.51.54,PhysRevLett.77.2154} as $\mathcal {D}^2+\mathcal {V}^2 \leq 1$, where $\mathcal {D}$ is the distinguishability of which-path information (a measure of particle-property), and $\mathcal {V}$ is the contrast visibility of interference (a measure of wave-property). 
In this regard, getting which-path information exclusively reduces the interference visibility, and vice versa. 
This double-path duality relation has been  tested in pioneer experiments~\cite{PhysRevLett.81.5705} and recently in delayed-choice measurements~\cite{PhysRevLett.100.220402,NiceDC}.  }
\end{quotation}
}]

\newpage 
\clearpage

\noindent 

Bohr's complementarity is one of the most central tenets of quantum physics~\cite{Bohr}. 
The paradoxical wave-particle duality of quantum matters and photons has been thoroughly tested in the famous Young's double-slit or Mach-Zehnder double-path interferometric systems, e.g. enabling the demonstrations of de Broglie's hypothesis and Wheeler's gedanken experiments \cite{Grangier_1986,RevModPhys.60.1067,Durr1998,C60,Zeilinger2005,Wootters79,Shadbolt2014,Wheeler,RevModPhys.88.015005,Jacques966,PhysRevLett.100.220402}. 
The quantum object  exhibit both the wave and particle nature~\cite{Wootters79}, but exclusively, depending on measurement apparatus where the choice of which-measurement it will take can be  delayed made in order to rule out too-naive interpretations of quantum complementarity~\cite{Wheeler,RevModPhys.88.015005,Jacques966,PhysRevLett.100.220402,USTCDC,NiceDC,BristolDC,NJUDC}. 
All experiments to date however have been implemented in the framework of double-path interference, while it is of both fundamental and technological interests to study quantum complementarity in an unexplored but general regime of sophisticated multipath interferometric implementations. 
Here we report an experimental demonstration of generalized multipath wave-particle duality in a quantum delayed-choice experiment, implemented by an on-chip multipath interferometer with a large-scale integration of silicon nanophotonics. 
Single photons propagating in the multipath system display sophisticated transitions between wave and particle characters -- distinct to those  in simple double-path systems, delayed determined by the choice of quantum-controlled measurements.   
We quantify the particle nature by multimode which-path information and  wave nature by multipath quantum coherence  directly probed from interference, and demonstrate the generalization of Bohr's multipath duality relation for genuine wave-particle quantum superposition, but not hold for its classical mixture counterpart. 
Our work provides deep insights into multidimensional quantum physics and benchmarks  high controllability and versatility  of large-scale integrated quantum nanophotonic technology.

\begin{figure*}[ht!]
\centering 
\includegraphics[width=0.995\textwidth]{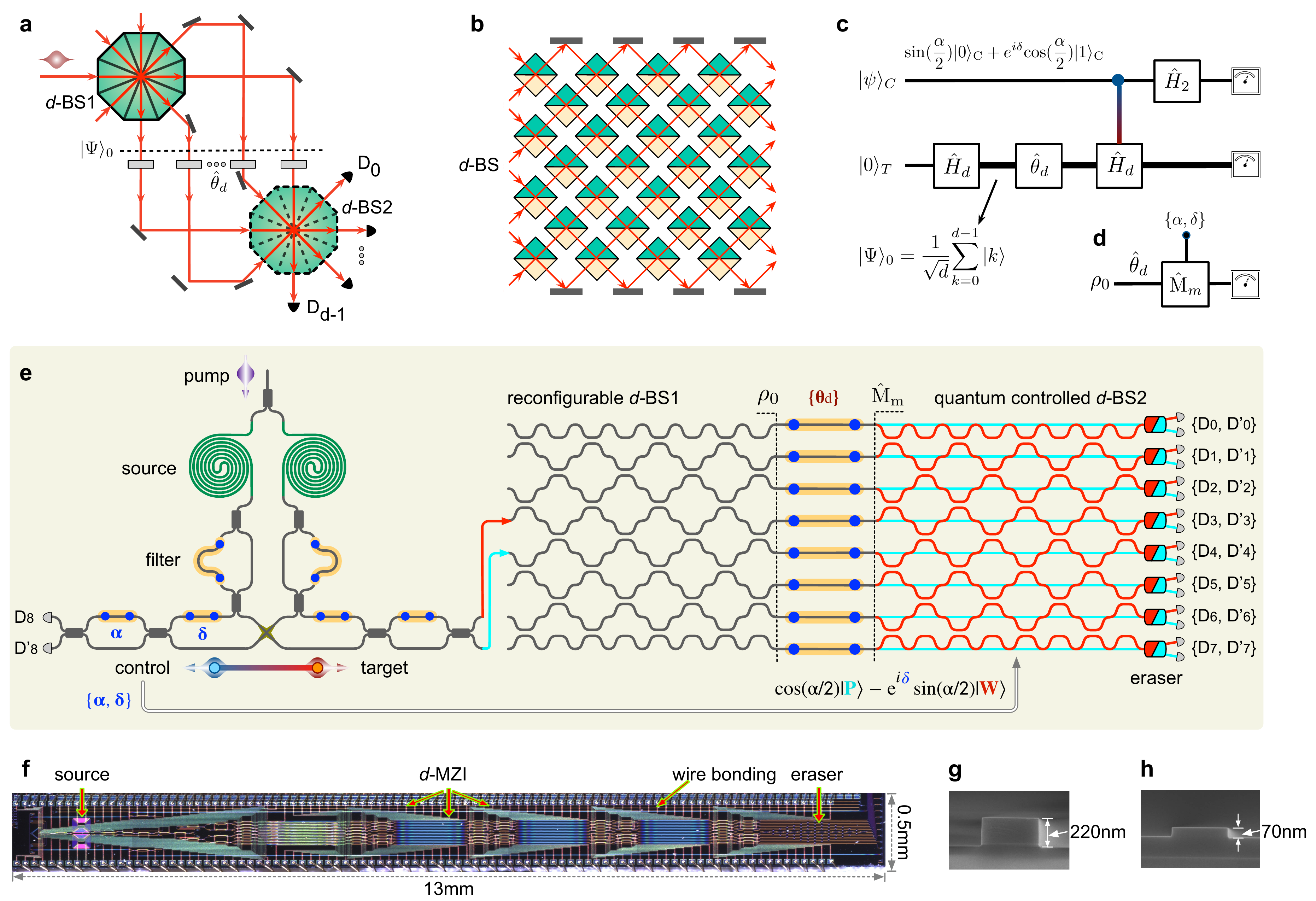}
\caption{\textbf{A quantum delayed-choice multipath experiment on a large-scale silicon-integrated quantum nanophotonic chip.} 
\textbf{a,} Diagram for a $d$-path MZI, including two $d$-mode beamsplitters ($d$-BS1, $d$-BS2), and a phase array \{$\theta_d$\} in the $d$ paths. The presence or absence of $d$-BS2 (dashed) allows the measurement of wave or particle nature in the delayed-choice manner.   
\textbf{b,} An example of $d$-BS by nesting conventional bulk-optic BSs. The $d$-BSs enable the implementation of general Hadamard operator $\hat{H}_d$.  
\textbf{c} and \textbf{d,}  Quantum circuit representation. The state of $\hat{H}_d$ operator is coherently  entangled with the control qubit $|\psi\rangle_C$. Our target is to quantify the wave-particle dual nature of a maximal coherent state $\rho_0=|\Psi\rangle_{00}\langle\Psi|$ by the delayed choice of which-measurement $\hat{\text{M}}_m$. The $\hat{\text{M}}_m$ represents the $d$-mode which-measurement operator, where $m$ is the measurement setting. 
\textbf{e,} Simplified schematic, and \textbf{f,} an optical microscope image of the device.  
A pair of path-entangled photons are generated via the spontaneous four-wave mixing (SFWM), by pumping  photon-pair sources using a continuous-wave laser at 1550nm. 
The target photon (red, signal) is sent through a $d$-path interferometer, undergoing either a wave (red) or particle  (cyan) process. 
A $d$-mode eraser then ensures quantum indistinguishability between the two processes. Which-measurement $\hat{\text{M}}_m$ the photon will take is decided by the state of $d$-BS2, crucially, that is coherently entangled with the control photon \{$\alpha, \delta$\} (blue, idler). Arbitrary \{$\theta_d, \alpha, \delta$\} can be chosen on the chip. 
The $d$-BSs  nested by 2-path MZIs are fully reconfigurable. 
The silicon chip monolithically integrates 95 phase shifters (yellow), 158 beamsplitters, 58 waveguide crossers, 42 grating couplers, and 2 SFWM sources, in total 355 components (some are not shown for clarity), among one of the largest quantum photonic devices. 
The single-photons are ultimately coupled off chip for detection by an array of superconducting nanowire single-photon detectors $\{\text{D}_0$,$\text{D}^{'}_0$,...$\text{D}_8$,$\text{D}^{'}_8\}$. 
\textbf{g} and \textbf{h,} Scanning electron microscope (SEM) images for the fabricated silicon nanophotonic waveguides with 220 nm full etch and 70nm shallow etch.  
}
\label{fig:device}
\end{figure*}

The interference of multiple states or processes  is a general wave phenomenon in all wave physics, such as optics, microwave, acoustics, solid-state and atomic physics.  
Since the birth of quantum mechanics, it has long been of fundamental interests to understand multipath interference of quantum mechanical wavefunctions and multidimensional quantum systems~\cite{PhysRev.47.777,Schrodinger1935,Lapkiewicz2011,SORKIN,Sinha418,Cottere1602478}. 
Although the dual nature of photons~\cite{Grangier_1986}, neutrons~\cite{RevModPhys.60.1067}, atoms~\cite{Durr1998}, and molecules~\cite{C60} has been revealed in conventional double-slit (or path) experiments, the quantum nature represented as Bohr's principles of complementarity and superposition remains ambiguous in multipath interferometric quantum systems~\cite{PhysRevA.51.54,Durr,Huber,Bera,PRLCoherence,PlenioCoherence}.  
Figure~\ref{fig:device}A sketches a multipath Mach-Zehnder interferometric delayed-choice experiment with single photon. 
The photons can either take all $d$ paths simultaneously, or  one of the $d$ paths, or  everything in between, delayed determined by the choice of $d$-mode which-measurement. 
In contrast to the double-slit implementation, the well-known duality relation initialized by Wootters~\cite{Wootters79} and formalized by Greenberger, Jaeger, and Englert~\cite{Greenberger,PhysRevA.51.54,PhysRevLett.77.2154}, well describes the wave-particle complementarity, however it cannot be simply generalized in the multipath  experiment~\cite{PhysRevLett.86.559,Durr,Huber,Bera,PRLCoherence}. 
%
There are several major open questions remaining: Can Bohr's duality relation still hold in the multipath interferometric experiment? 
Are there any good measures of multipath wave and multimode particle properties that are  accessible in experiment? 
Does  single photons preserve the inherent dual nature in the multipath delayed-choice scenario? 
Revealing the unknowns are essential to understand multimode quantum superposition and quantization in complex quantum systems. 
Apart from fundamental interests, the characterization of mutilmode quantum properties in controllable systems may provide the ground of developing  multidimensional quantum technologies. 
For example, quantifying mutilmode coherence from sophisticated multipath interference patterns is of practical significance~\cite{Qureshicoherence}, in the light of recent reassessment of coherence as a key resource in quantum information~\cite{coherencereview,PlenioCoherence}, while it has always been a core concept underlying the theory of quantum mechanics. 
In general, when single photons pass in multiple paths, a superposition of multiple modes naturally forms a \textit{qudit} state. 
Promising prospects of {qudit}-based quantum applications have been well acknowledged, such as  
noise-robust multidimensional entanglement~\cite{Wang16D,Friis2019}, resource-efficient quantum computations and simulations~\cite{Reimer2019,Superconducingqudit}, and high-capacity quantum communications~\cite{Cerf2002,Ecker_2019}; however, the deep understanding of the most elementary physics of multidimensional quantum systems is highly demanded. 
Any explorations of multidimensional quantum science and technology strongly rely on the quantum platform that can be operated with high levels of controllability, efficiency and versatility~\cite{Erhard2020}. 
It is here  the integrated-optics implementation provides one of the most competitive multidimensional quantum platforms~\cite{Wangreview}.

Here, we report a quantum delayed-choice multipath experiment and demonstrate a generalization of wave-particle duality relation. 
The wave-particle nature of single photons propagating in a $d$-path ($d$ up to 8) interferometer is observed, that is delayed determined by the state of a $d$-mode quantum-controlled beamsplitter. 
Qualitative wave-particle transitions having a high fidelity of $0.99$ between theoretical and experimental results, and quantitative multipath duality relations are both confirmed in the context of delayed-choice. 
We demonstrate that quantum coherence is a good measure of the wave-property in $d$-path interference, and the amount of coherence can be directly probed from interference patterns, without accessing the  density matrix. 
The $d$-mode which-path information is identified, and quantum randomness is efficiently generated. 
All demonstrations are enabled by realizing a multipath delayed-choice interferometric system on a large-scale silicon nanophotonic quantum chip that integrates 355 optical components and 95 shifter-shifters. 

Figure \ref{fig:device}A shows a diagram of $d$-path Mach-Zehnder interferometer ($d$-MZI) consisting of \textit{d} arms and two $d$-mode beamsplitters ($d$-BSs). 
An array of individually reconfigurable phases of \{$\theta_k\}^{d-1}_{k=0}$ are applied on the $d$-path between $d$-BS1 and $d$-BS2. To implement the $d$-BSs, a general scheme is to nest  $(d^2-d)/2$ standard 2-BSs 
(see a bulk-optic example in Fig.\ref{fig:device}B).   
For simplicity, we consider the case that $d$-BSs are balanced, thus the state emerging from the $d$-BS1 is a maximally coherent state as $\rho_0=\ket \Psi_{00}\bra\Psi$ and $\ket{\Psi}_0=\frac{1}{\sqrt{d}} \sum_ {k=0}^{d-1} {\ket{k}}$, where $\{\ket{k}\}^{d-1}_{k=0}$ is the logical basis  that defines the reference frame. 
The state of $d$-BS2 represents which-measurement the photon will confront. 
When the $d$-BS2 is inserted (removed), the $d$-MZI is closed (open), the detected probabilities at $\{\text{D}_0,...\text{D}_{d-1}\}$ are dependent (independent) on the \{$\theta_k\}^{d-1}_{k=0}$ configuration, thus revealing the $d$-path wave-like interference ($d$-mode particle-like quantization). 
In order to probe the genuine duality, the state of $d$-BS2 has to be determined after the photon has entered the $d$-MZI. 
In Fig.\ref{fig:device}C, we suggest a modified version of quantum-controlled delayed-choice that was recently proposed by Ionicioiu and Terno~\cite{QBStheory} and implemented in double-path experiments~\cite{USTCDC,NiceDC,BristolDC,NJUDC}. 
In our multipath quantum delayed-choice scheme, the state of a general $d$-dimensional Hadamard operator $\hat{H}_d$ is coherently entangled with the control qubit  $|\psi\rangle_C$. 
The Hadamard $\hat{H}_d$ is implemented by a balanced $d$-BS, with the element of $h^{(d)}_{i,j}=\frac{1}{\sqrt{d}}(-1)^{i\odot j}$, where $i \odot j$ is the bitwise dot product of the binary representations of $i$ and $ j$. 
Note that realizing the multidimensional quantum-controlled $d$-BS2 is an essential, yet challenging task. We  adopt an approach in which the state can be seen as entangled with the process in our experiment. 

We devise a large-scale silicon-integrated quantum nanophotonic device for the implementation of the delayed-choice $d$-path interferometric experiment, that features high phase stability and scalability~\cite{Wangreview}. 
Figure~\ref{fig:device}E illustrates a simplified diagram of the device to implement the circuit in Fig.\ref{fig:device}C. 
Our task is to test the wave-particle dual nature of the target photon $\rho_0$ by choosing which-measurement $\hat{\text{M}}_\text{m}$ (Fig.\ref{fig:device}D). We remark that which-measurement is delayed determined by the state of quantum $d$-BS2, that is entangled with the control photon.  

The device includes four parts: an entangled photon-pair source, a $d$-path MZI, a quantum-controlled $d$-BS, and a $d$-mode  eraser. 
We first prepare a maximally entangled state $(\ket{0}_C\ket{0}_T+\ket{1}_C\ket{1}_T)/\sqrt{2}$ in two integrated spontaneous four-wave mixing (SFWM) sources~\cite{Wang16D}, where $\ket{0}_{C,T}$ and $\ket{1}_{C,T}$ are path-encoded logical states of the control and target photons. 
The target photon undergoes a $\hat{H}_d$ transformation by the $d$-BS1, and a phase operator $\hat{\theta}_d$ by  the phase array. 
Depending on the control state of  $\ket{0}_C$ or  $\ket{1}_C$, the target photon coherently evolves either as a wave  or particle, respectively, resulting in a state-process entanglement: 
\begin{equation}
\frac{1}{\sqrt{2}}{(\ket{0}_C\ket{\text{P}}_T+\ket{1}_C\ket{\text{W}}_T)},
\end{equation}

\vspace{-4mm}
\begin{equation}
\begin{split}
\ket{\text{P}}_T=\frac{1}{\sqrt{d}}&\sum_{m=0}^{d-1} e^{i\theta_m}\ket{m}_T, 
\quad
\ket{\text{W}}_T=\frac{1}{\sqrt{d}}\sum_{m=0}^{d-1} \sum_{k=0}^{d-1} h^{(d)}_{mk}e^{i\theta_k}\ket{m}_T, 
\label{eq:state}
\end{split}
\end{equation}
where $ \ket{\text{P}}$, $\ket{\text{W}}$ represent the states taking the particle and wave processes, respectively. The which-process information is erased at a $d$-mode quantum eraser (see Fig.\ref{fig:device}E), ensuring quantum mechanical indistinguishability between the two processes. 
See Supplementary Secs.\ref{sec:Device},\ref{sec:DCscheme} for more details. 
This state-process entanglement approach has  been  adopted for implementing double-path delayed-choice experiments~\cite{NiceDC} and for quantum simulations~\cite{ Wang:QHL}.

We operate the control photon state as $\sin\frac{\alpha}{2}\ket{0}_C+e^{i\delta}\cos\frac{\alpha}{2}\ket{1}_C$, where \{$\alpha$, $\delta$\} represents the \{$\sigma_y$, $\sigma_z$\} rotations, and selected the events when the detector $\text{D}_{8}^{'}$ clicked (see Fig.\ref{fig:device}E). 
Owing to the presence of entanglement between the control photon and the quantum state of $d$-BS2, the $d$-BS2 is thus in a superposition of presence and absence ($\cos\frac{\alpha}{2}\hat{I}-i {e}^{i\delta}\sin\frac{\alpha}{2}\hat{H}_d$). 
Note that the first term denotes the measurement of particle character, while the second term denotes the measurement of wave character. 
Figure~\ref{fig:device}D represents the framework of delayed choice of which-measurement  $\hat{\text{M}}_m=\ket{m}\bra{m}$ performing on the target photon $\rho_0$,  where $\ket{m}$ forms the wave-particle measurement basis: 
\begin{equation}
\ket{m}= \frac{1}{\sqrt{N_{d}}}\sum^{d-1}_{k=0}{ \left(\Delta_{(m-k)} \cos \frac{\alpha}{2}+{i } \frac{e^{-i\delta}}{\sqrt{d}}(-1)^{m\odot k} \sin \frac{\alpha}{2}\right) \ket{k}}, 
\label{eq:measurement}
\end{equation}
where $m={0...,d-1}$ is measurement settings; $\Delta_{x}$ is the Kronecker delta function; $N_{d}$ is a normalization coefficient. See details in Supplementary Sec.\ref{sec:duality}. 
The probability of obtaining the $m$-th measurement outcome is quantified by $\text{Tr}{[\hat{\text{M}}_m\rho']}$, where $\rho'$ is the state after the $\hat{\theta}_d$. 
In our experiment, the probabilities for each measurement are calculated by the normalization of two-fold coincidences over the $d$-ports.

The choice of  which-measurement  $\hat{\text{M}}_m$ -- delayed determined by the  \{$\alpha, \delta$\} state of the control photon, allows us to observe the $d$-path wave-particle transition and to test the $d$-path duality relation of the target photon.  
In Eq.\ref{eq:measurement},  $\alpha$ refers to the amplitude of wave and particle properties, and the inherent $\delta$ phase identifies the genuine quantum particle-wave superposition. 
If $\alpha=0$, $d$-BS2 is in the off-state and the $\hat{\text{M}}_m$ discloses the particle-nature. 
Hence, the photon registers each of the detectors (Fig.\ref{fig:device}E) with a probability of $1/d$, leading to the observation of $d$-mode quantized distributions. 
If $\alpha=\pi$, $d$-BS2 is in the on-state and $\hat{\text{M}}_m$ reveals the wave-nature. 
In this regard, the probability of detecting the photon relies on $\{\theta_d\}$, building up $d$-path wave interference patterns.  
If $0<\alpha<\pi$, $d$-BS2 is in a superposition of the on- and off-state and it thus allows the observation of wave-particle nature simultaneously. 
We remark that the choice of $\hat{\text{M}}_m$ and the dual property of the target photon remain undetermined, until the control photon has been detected. 
This is because of the essence of entanglement that information is non-locally shared between the two photons.

Our quantum chip is designed for $d$-path ($d\le8$) experiments, in which  the number of paths and  mode number of $d$-BSs can be reconfigured. 
The $d$-BSs are formed by a squared mesh of 2-BSs (each is a 2-path MZI  for full reconfigurability). 
The chip integrates 95 phase shifters that are individually addressed and electronically driven. 
A telecom-band  laser was used to generate two entangled photons under the SFWM process. 
The signal photon is considered as the target photon, while the idler photon plays the role of control photon. 
The two photons were routed off the chip for detection by superconducting nanowire single-photon detectors $\{\text{D}_i, \text{D}_i^{'}\}$, $i$=0,...8 (see Fig.\ref{fig:device}E). The fabricated devices and waveguides are shown in Figs.\ref{fig:device}F-H.  See Supplementary Sec.\ref{sec:Device} for more details of the device and setup.

\begin{figure}[t]
\centering 
\includegraphics[width=0.27\textwidth]{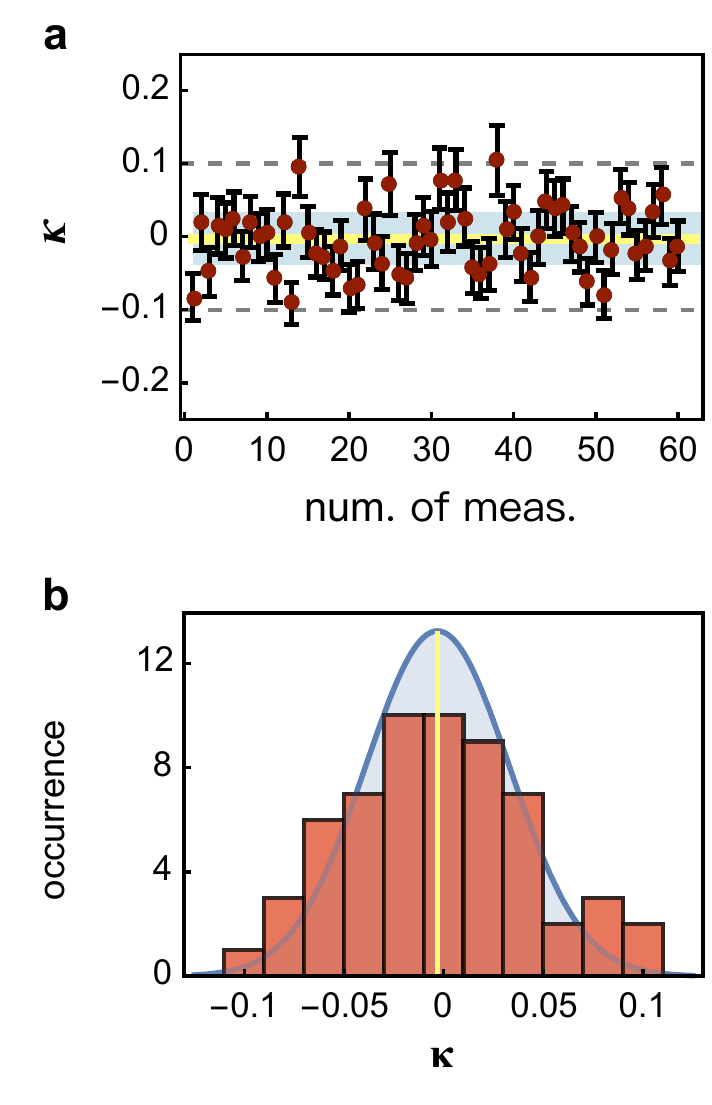}
\caption{\textbf{Measurement of high-order interference in the $d$-path interferometer.}
\textbf{{a,}} Measured $\kappa$ values for the fourth-order interference that is normalized to the second-order interference. In total, 60 measurements (red points) are performed independently. 
Error bars ($\pm\sigma$) are estimated from photon Poissonian statistics. Yellow line denotes the mean value of $\kappa$, and shaded regime shows one standard deviation for all $\kappa$. 
\textbf{{b,}} Histogram of all measured $\kappa$ values.  Shaded regime shows a fitted Gaussian profile distribution.     
All  data are measured at the prime maxima of the complete wave interference fringes. 
The measured tight bound of $\kappa=-0.0031\pm0.0047$ allows us to rule out the presence of high-order interference. 
}
\label{fig:HighOrders}
\end{figure}

\begin{figure*}[ht!]
\centering 
\includegraphics[width=0.90\textwidth]{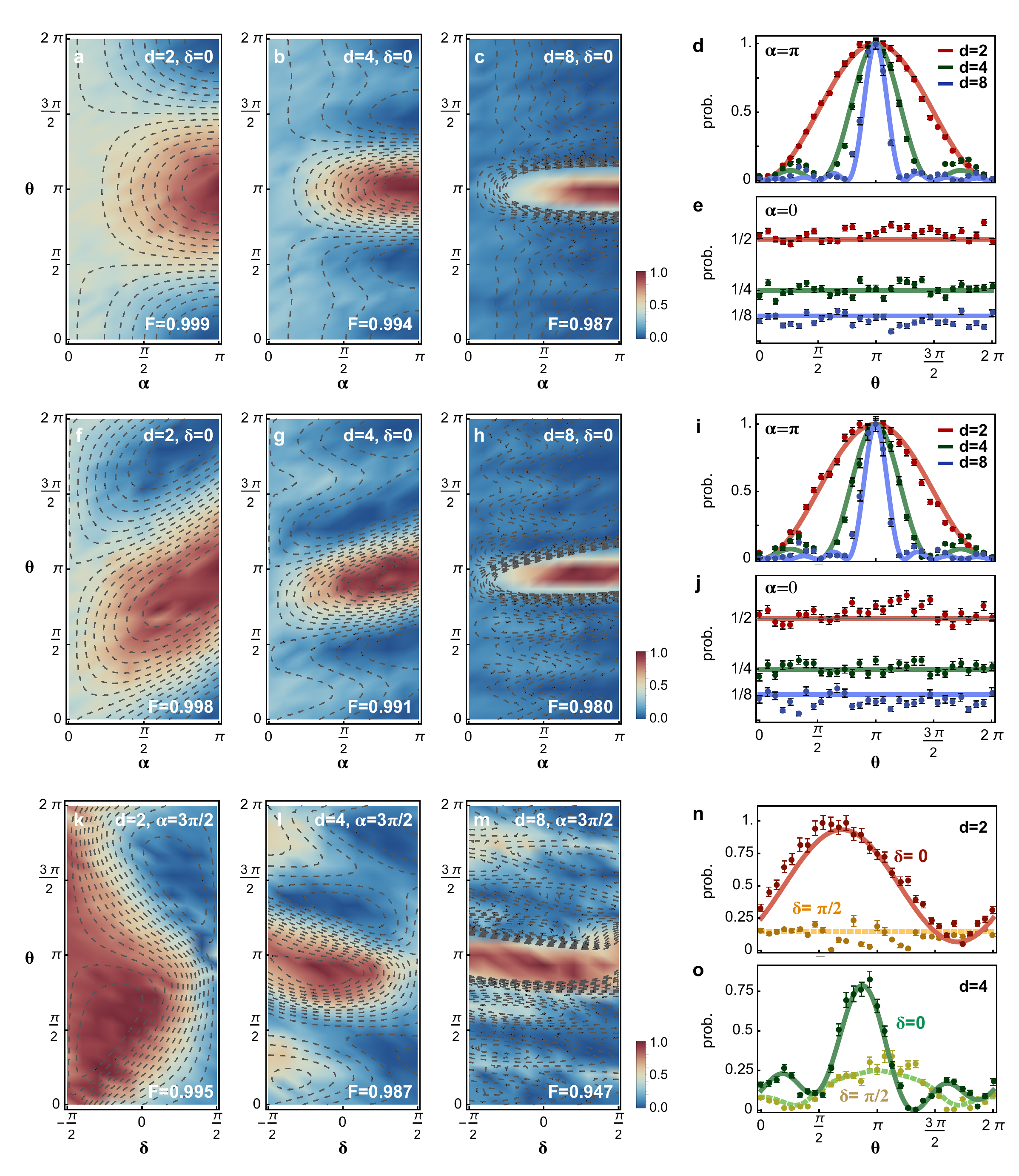}
\caption{\textbf{Experimental observations of multipath wave-particle transition in the delayed-choice experiment.} 
Measured transitions between particle and wave properties in several different scenarios: \textbf{a}-\textbf{c,} classical-mixture;  \textbf{f}-\textbf{h,} quantum-superposition; and \textbf{k}-\textbf{m,} intrinsic coherent quantum-superposition. 
They are quantified by probability distributions (normalized coincidences) for different \{$\alpha,\delta$\} of the control and  ${\theta}_d$ of the target (${\theta}_d=k(\theta-\pi)$ was chosen), in the 2, 4 and 8-path experiments. 
Density distributions (colored) represent experimental data, while contour lines (dashed) represent theoretical results. The ${F}$ denotes the classical fidelity $\sum_i \sqrt[]{p_iq_i}$, where $p_i$ and $q_i$ are the  measured and theoretical probabilities, respectively. High fidelities are obtained for all measurements. 
Results in \textbf{(a}-\textbf{c)} are consistent with classical optical multi-slit interference.  
The asymmetry of transition patterns in the quantum case (\textbf{f}-\textbf{h}) stem from the quantum interference between wave and particle properties.  
The $\delta$-dependence of interference patterns in \textbf{(k}-\textbf{m)}, measuring at $\alpha={3\pi}/{2}$, further confirms the existence of genuine wave-particle superposition. 
\textbf{d}-\textbf{e,} Classical fringes, and \textbf{i}-\textbf{j,} quantum fringes for the full-wave case at $\alpha=\pi$, and full-particle case at $\alpha=0$, for $d=$2, 4 and 8. 
The interference fringe becomes sharper for $d$-path interference; the multimode quantization results in a $1/d$ probability at the outport. 
Only when $\alpha=\{0,\pi\}$,  classical and quantum fringes agree with each other. 
Examples of $\delta$-dependence interference fringes for $\delta=\{0, {\pi}/{2}$\} and $\alpha={3\pi}/{2}$  in the \textbf{(n),}  2-path, and \textbf{(o),} 4-path experiments. 
The construction or destruction of interference appears by controlling the internal $\delta$ phase. 
Points represent experimental data, while lines represent theoretical values. All error bars ($\pm 3\sigma$) are estimated from photon Poissonian statistics. 
}
\label{fig:duality}
\end{figure*}

Prior to  testing the wave-particle duality, we first rule out the existence of high-order interference in the $d$-path experiments. 
Bohr's rule is based on the postulate that quantum interference only consists of mutual coherence terms taken over all pairs of paths. As an example, we implemented 4-path interference. 
We measured the normalized Sorkin parameter $\kappa$, a ratio of high-order interference to second-order interference~\cite{SORKIN,Sinha418,Cottere1602478}, and we obtained the tight bound of $-0.0031\pm0.0047$  for the fourth-order interference (see Fig.\ref{fig:HighOrders}). Our experimental results confirm the absence of high-order interference, within an accuracy of $10^{-3}$ of the bound. Main experimental errors come from the non-perfect closure of certain paths. 
See Supplementary Sec.\ref{sec:highorder} for more measurement details.

Figure~\ref{fig:duality} reports the experimental results for  $d$-path wave-particle transitions. Here, probability distributions for the $\hat{\text{M}}_0$-measurement are plotted, corresponding to the detection of the target photon at the ports $\{\text{D}_0, \text{D}_0^{'}\}$. 
In Fig.\ref{fig:allotherports} it shows the $m$-measurement results for other ports $\{\text{D}_i\}^{d-1}_{i=1}$. 
In our measurement, the phase ${\theta}_k=k(\theta-\pi)$ was chosen, where $\theta$ $\in$ $[0, 2\pi]$, but an arbitrary ${\theta}_k$ phase can be set. 
We obtained the continuous transition of $d$-path  duality, from the full-particle nature at $\alpha=0$ to the full-wave nature at $\alpha=\pi$, in two different scenarios: classical-mixture (Figs.\ref{fig:duality}A-E) and quantum-superposition (Figs.\ref{fig:duality}F-J). 
Classical  wave-particle mixture refers to the state  $(\cos^2\frac{\alpha}{2}\ket{\text{P}} \bra{\text{P}}+\sin^2\frac{\alpha}{2}\ket{\text{W}} \bra{\text{W}})$, while quantum wave-particle  superposition refers to the state  $\frac{1}{\sqrt{N}}(\cos\frac{\alpha}{2}\ket{\text{P}}- i e^{i\delta}\sin\frac{\alpha}{2}\ket{\text{W}})$, where $N$ is a normalization coefficient, representing the ultimate states after the quantum erasure. 
Details are given in Supplementary Sec.\ref{sec:DCscheme}. 

In the case of  $2$-path classical mixture, Fig.\ref{fig:duality}A shows a sinusoidal interference fringe at $\alpha=\pi$ representing the full-wave nature, and  the detection probability approaches nearly $1/2$ at $\alpha=0$ representing the full-particle nature. The observation of 2-path wave-particle transition  is consistent with the results  in~\cite{NiceDC,BristolDC,NJUDC}. 
In contrast, in the $d$-path experiments  (Figs.\ref{fig:duality}B and \ref{fig:duality}C), at $\alpha=\pi$ we observed interference patterns that feature sharper distributions with an increment of $d$, confirming the $d$-path wave nature; at $\alpha=0$ we observed a $1/4$ ($1/8$) probability for $d=4$ ($8$), confirming the $d$-mode particle nature;  for $0<\alpha<\pi$,  intermediate particle-wave  behaviors  were revealed.  
The results for $\alpha=\{0,\pi\}$ are replotted in Figs.\ref{fig:duality}D, E, which are expected in classical optical multi-slit interference.  

We now report the unique feature of quantum wave-particle superposition in $d$-path experiments. 
In Figs.\ref{fig:duality}F-H, the probability distributions represent asymmetry with respect to $\theta=\pi$, while the distributions for classical mixture in Figs.\ref{fig:duality}A-C remain symmetric. The asymmetry is because of the quantum interference between wave and particle properties (see Eq.\ref{eq:quantumPro}). 
Only when choosing the full-particle ($\alpha=0$) or full-wave ($\alpha=\pi$) point, the distributions for classical (Figs.\ref{fig:duality}D, E) and quantum cases (Figs.\ref{fig:duality}I, J) are in agreement.
 When $\alpha \neq \{0, \pi\}$, the quantum distributions are remarkably distinct from the classical ones. 
Quantum distributions however tend to be less asymmetric for high $d$-path interference, becoming more classical (see analysis in Sec.\ref{sec:larged}). 
Figures \ref{fig:duality}K-M show 
quantum interference of multipath wave and multimode particle properties regarding the inherent phase $\delta$. 
We set $\alpha =3\pi/2$ that corresponds to the maximal wave-particle superposition. 
The  $\delta$-dependence of  distributions confirms the genuine quantum wave-particle superposition, while $\delta$-variation is absent in the case of classical mixture (see  the explicit forms in Eqs.\ref{eq:quantumPro},  \ref{eq:classicalPro}). 
It is notable that by controlling the $\delta$ phase, the quantum interference of the wave and particle properties can be steered. 
For example, in the 2-path case, Fig.\ref{fig:duality}N shows constructive interference for $\delta=0$ and destructive interference for $\delta=\{\frac{\pi}{2}$, $ -\frac{\pi}{2}\}$. 
In the case of $4$-path interference (see Fig.\ref{fig:duality}O), more wave-like characters appear when $\delta$ is set as $\frac{\pi}{2}$.

\begin{figure*}[ht!]
\centering 
\includegraphics[width=0.55\textwidth]{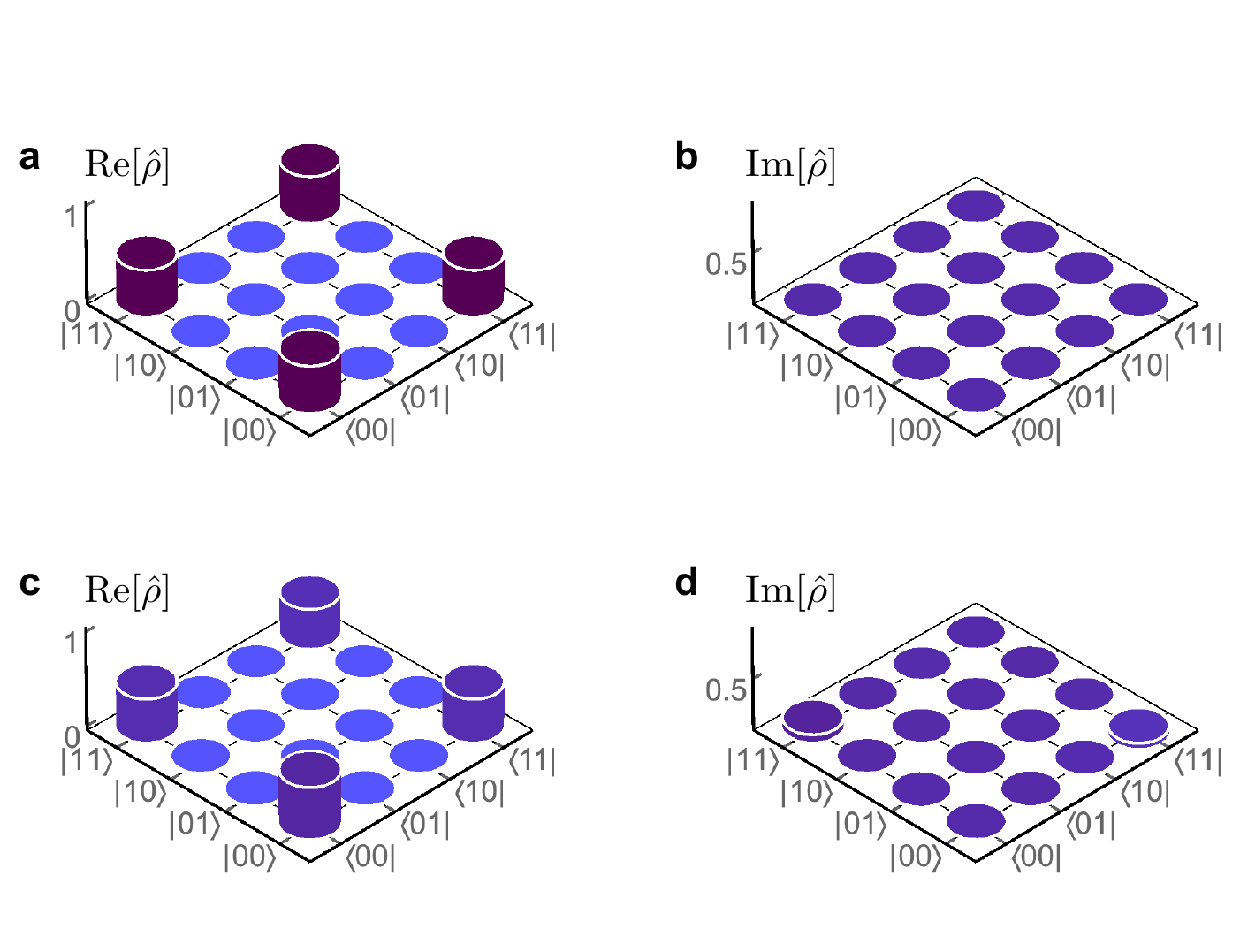}
\caption{\textbf{| Experimental quantum state tomography.}
\textbf{a,} Real and \textbf{b,} imaginary part of the ideal maximally entangled state  $(\ket{00}+\ket{11})/\sqrt{2}$. \textbf{c,} Real and \textbf{d,}  imaginary part of the experimentally reconstructed density matrix. In our experiment, over-complete tomography techniques with 36 measurements in total were used to reconstruct the state. Quantum state fidelity of $F_Q= 0.962\pm0.002$ was measured, which is defined as $\bra{\phi}\hat{\rho}\ket{\phi}$, where $\hat{\rho}$ represents the reconstructed state in \textbf{(c} and \textbf{d)} and $\ket{\phi}$ is the ideal  Bell state in \textbf{(a}  and \textbf{b)}. 
}
\label{fig:rho}
\end{figure*}

All measurements in Fig.\ref{fig:duality} were performed in the computational basis \{$\ket{0}$, $\ket{1}$\}, whch are well in agreement with theoretical predictions. 
We obtained the classical fidelities (see definition in Fig.\ref{fig:duality}'s caption) 
with a mean  of $0.998\pm 0.001$ for 2$d$-, $0.991\pm 0.003$ for 4$d$-, and $0.980\pm 0.007$ for 8$d$-path experiments, respectively. 
Since entanglement is playing an enabling role in the delayed-choice measurement of $d$-path wave-particle duality, we repeated the experiments in the complementary basis \{$\ket{+}$, $\ket{-}$\}, where $\ket{\pm}$ denotes $(\ket{0} \pm\ket{1})/\sqrt{2}$. 
We again obtained coherent wave-particle transitions with high fidelities in the complementary basis, as shown in Fig.\ref{fig:Xbasis}. 
Moreover, to exclude the presence of local hidden variables that may contribute the delayed-choice measurement, we verified entanglement by both performing quantum state tomography of the entangled state, and  demonstrating the violation of the Bell-CHSH (Clauser-Horne-Shimony-Holt) type inequality~\cite{PhysRevLett.23.880}. 
The quantum state fidelity of $0.962\pm 0.002$ was obtained (see Fig.\ref{fig:rho}). 
We also measured the Bell value of $2.75\pm 0.04$, which violates the classical bound by $18.8 \sigma$, confirming the existence of strong entanglement through the device.

We next report experimental results of a generalized multipath  duality relation in the delayed-choice interferometer. 
It is of fundamental interest to develop a general framework to describe the multipath duality  and to quantify  wave and particle properties~\cite{PhysRevA.51.54,Bera,Durr,Huber,PRLCoherence}. 
The conventional visibility, defined as  the contrast of interference fringe, fails to be a good wave measure for $d$-path ($d>2$) interference~\cite{PhysRevLett.86.559}; however, quantum coherence is now believed to be a good quantifier~\cite{Bera,Durr,Huber,PRLCoherence}. 
We adopt the $l_1$-norm coherence ($\widetilde{\mathcal{C}}_{l1}=\sum_{i \neq j} |\rho_{i j}|$) proposed by Streltsov \textit{et. al.} as a wave measure~\cite{PlenioCoherence}. 
Moreover, the capability of distinguishing which-path the photon has taken represents the $d$-path distinguishability. 
The formation of path-distinguishability for the 2-path case can be generalized to the $d$-path case~\cite{Durr,Huber,QureshiD}. 
The normalized coherence $\mathcal{C}_d$ and path-distinguishability $\mathcal{D}_{d}$ that we have used are given as: 

\begin{equation}
\mathcal{C}_d = \frac{1}{d-1} \sum\limits_{i \neq j}\left|\rho_{i j}\right|, ~~\mathcal{V}_d=\dfrac{1}{d-1}\dfrac{I_{max}-I_{inc}}{I_{inc}}, 
\label{eq:C}
\end{equation}

\begin{equation}
 \mathcal {D}_{d}=\sqrt{1-\left(\frac{1}{d-1} \sum\limits_{i \neq j} \sqrt{\rho_{ii}\rho_{jj}}\right)^{2}}, 
\label{eq:D}
\end{equation}
where $\rho$ represents the state for entire system having the target photon and measurement, as photon displays particle or wave nature is  dependent on the measurement apparatus. 
The off-diagonal elements $\rho_{ij}$ determine the $d$-path wave interference, while the diagonal elements $\rho_{ii}$ determine the distinguishability of $d$-mode path-information. 
The explicit forms of $\rho$ for $d$-path classical and quantum experiments are given in Supplementary Sec.\ref{sec:duality}. 
Bohr’s multipath duality rule is thus quantitatively generalized  as~\cite{QureshiD}:  
\begin{equation}
\mathcal {C}^2_{d} + \mathcal {D}^2_{d}  \leqslant  1, 
\label{eq:inequality}
\end{equation}
which saturates for pure state, $\textit{i.e.}$ wave-particle superposition state. See  proof in Supplementary Sec.\ref{sec:duality}.  
The definitions of  $\mathcal{C}_d$ and $\mathcal{D}_{d}$ well meet Dürr’s criteria~\cite{Durr}, and importantly, they represent macrovariables that capture the global features of the $d$-path interferometric system. Note when $d=2$, the generalized duality relation in Eq.\ref{eq:inequality} reduces to the fundamental statement by Jaeger \textit{et. al.}~\cite{PhysRevA.51.54}. 

\begin{figure*}[ht!]
\centering 
\includegraphics[width=1\textwidth]{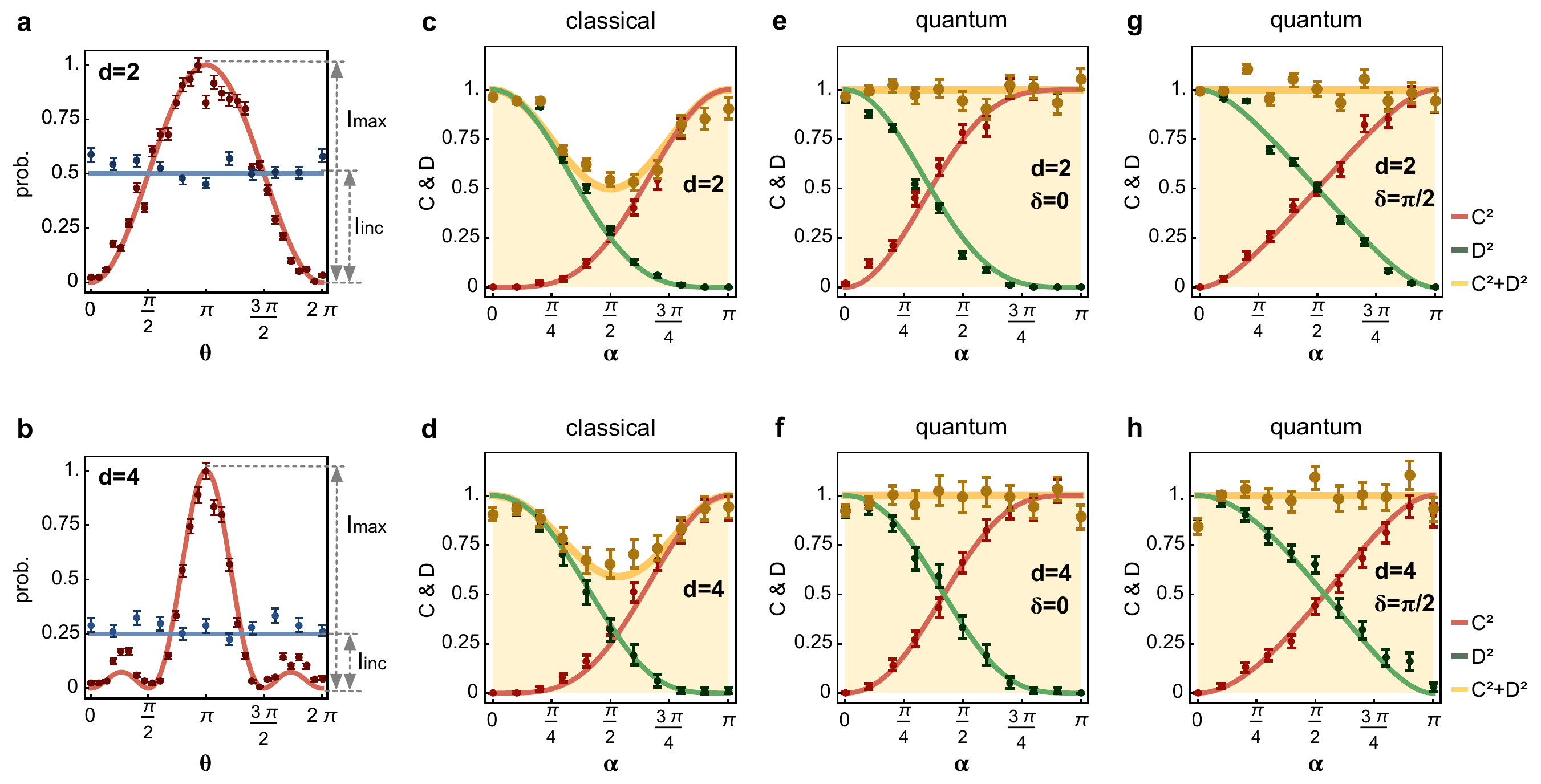}
\caption{\textbf{Experimental results of generalized multipath wave-particle duality relation in the delayed-choice experiment.} 
Measurement of generalized visibility ($\mathcal {V}_{d}$)  that is equivalent to the normalized $l_1$-norm coherence ($\mathcal {C}_{d}$) for \textbf{(a),} 2-path and \textbf{(b),} 4-path full-wave interference ($\alpha=\pi$).  
The $\mathcal {V}_{d}$ is determined by the disparity of the primary maxima $I_\text{max}$ and incoherent term $I_\text{inc}$, both directly read out from the $d$-path interference fringes, with no need to probe the full density matrix.   
Measurement of generalized  duality relation in the classical-mixture scenario: \textbf{c,} 2-path and \textbf{d,} 4-path experiments; in the genuine quantum-superposition scenario: \textbf{e,} 2-path $\delta=0$; \textbf{f,} 4-path $\delta=0$;  \textbf{g,} 2-path $\delta=\pi/2$; and \textbf{h,} 4-path $\delta=\pi/2$ experiments.   
The path-information $\mathcal {D}_{d}$ is meaured in a posteriori way, by the delayed choice of which-measurement $\hat{\text{M}}_m$. 
The bound of $\mathcal {C}^2_{d} + \mathcal {D}^2_{d}$ values are indicated within the orange-colored regime. The generalization of duality relation $\mathcal {V}^2_{d} + \mathcal {D}^2_{d} \leq 1$  is thus confirmed  in the $d$-path quantum-superposition experiment for different \{$d, \alpha,\delta$\} configurations \textbf{(e}-\textbf{h)}. 
In the classical mixture case \textbf{(c}, \textbf{d)}, the duality relation only hold at the full-wave and full-particle points at $\alpha=\{0,\pi\}$, independent on the setting of $\delta$. 
Points represent experimental data, while lines represent theoretical values. Error bars ($\pm \sigma$) are estimated from photon Poissonian statistics. 
}
\label{fig:inequality}
\end{figure*}

\begin{figure}[t!]
\centering 
\includegraphics[width=0.24\textwidth]{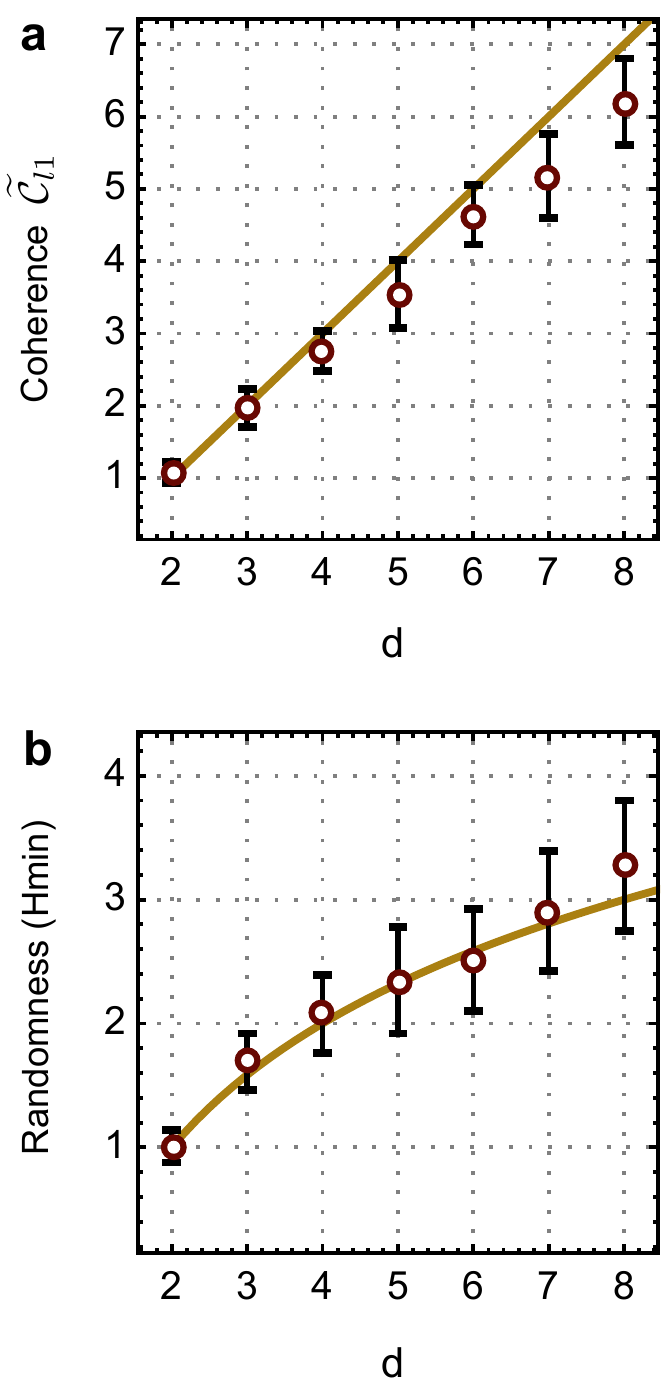}
\caption{\textbf{Characterization of multimode coherence and multimode quantization.} 
\textbf{a,} Quantification of the $l_1$-norm quantum coherence $\widetilde{\mathcal{C}}_{l1}$ from  the $d$-path full-wave interference patterns when $\alpha$ is chosen as $\pi$. 
The $\widetilde{\mathcal{C}}_{l1}  $ values are estimated from  the visibility by $\mathcal{V}_d (d-1)$, that are directly probed from the $d$-path interference patterns, without accessing the full density matrix. 
\textbf{b,} Measurement of quantum randomness entropy $H_{min}$ from the $d$-mode full-particle quantized distributions, when $\alpha$ is chosen as zero. 
More than one bit of randomness is obtained. The results in \textbf{(a} and \textbf{b)} are collected in the quantum wave-particle superposition measurement. 
Red points represent experimental data, while orange lines represent theoretical results. Error bars ($\pm \sigma$) are estimated from photon Poissonian statistics. 
}
\label{fig:ArbitraryD}
\end{figure}

The $\mathcal {C}_{d} $ and $\mathcal {D}_{d} $  were measured in the context of delayed choice of which-measurement $\hat{\text{M}}_0$ under the  control of  $\{\alpha,\delta\}$. 
It is notable that the measurement of which-path information $\mathcal {D}_{d}$ in our experiment is a \textit{posteriori} one~\cite{PhysRevLett.100.220402,NiceDC}. 
This is because of the fact that  $\hat{\text{M}}_0$ enables the projection into the particle-wave superposition basis,  that is delayed controlled from the full-wave to full-particle measurements. 
%
In order to measure the $\mathcal {C}_{d}$, we here adopt a newly derived general visibility $\mathcal {V}_{d}$ proposed by Roy and Qureshi~\cite{Qureshicoherence} (see definition in Eq.\ref{eq:C}). 
Importantly, $\mathcal {V}_{d} $ is exactly equivalent to the normalized coherence, \textit{i.e.}, $\mathcal {C}_{d} =\mathcal {V}_{d} $, and it can be directly read out from interference patterns characterized by \{$I_{max}$, $I_{inc}$\} (see Figs.\ref{fig:inequality}A,B), without  knowledge of the density matrix of the system. 
We obtained the primary maxima $I_{max}$ of the interference fringes, and the incoherence contribution $I_{inc} = \frac{1}{d}\sum_{i=0}^{d-1} \rho_{ii}$. Take the case for $\alpha=\pi$ as examples, Figs.\ref{fig:inequality}A and B show the interference patterns for the 2-path and 4-path cases, from which the \{$I_{max}$, $I_{inc}$\} values and thus the $\mathcal{V}_d$ values can be directly obtained (results for $d \in [2, 8]$ are reported in Fig.\ref{fig:FigFringeAllD}). 
To quantify which-path information $\mathcal {D}_{d} $, we measured the diagonal elements $\rho_{ii}$, by recording the probability of $i$-outcomes only having the ${i}$-path open. 
We measured $\mathcal{V}_d$ and $\mathcal{D}_d$ for different $\{\alpha,\delta,d\}$.  
Figures \ref{fig:inequality}C,E (D,F) report the experimental results for the 2-path (4-path) case, when $\delta$ is set to be $0$. 
The results for the $ \delta $ of $\frac{\pi}{2}$ are shown in  Figs.\ref{fig:inequality}G,H.

Figures \ref{fig:inequality}C-H report the experimental results of the generalized multipath duality relation for  wave-particle quantum-superposition and classical-mixture cases. 
In the genuine wave-particle quantum-superposition case, in  Figs.\ref{fig:inequality}E-H,  we demonstrate that the generalized duality relation holds the tight bound of unity as $\mathcal {C}^2_{d} + \mathcal {D}^2_{d} = 1$ in different setting of \{$d, \alpha, \delta$\}. 
This is however not true for the classical-mixture case, where one cannot reach the upper bound and the equality does not saturate at $\alpha \neq \{0, \pi\}$, resulting in $\mathcal {C}^2_{d} + \mathcal {D}^2_{d} < 1$. 
The bound can only be approximated at $\alpha=\{0,\pi\}$ for full-particle or full-wave states, as shown in Figs.\ref{fig:inequality}C,D. 
The quantitative results in Fig.\ref{fig:inequality} are consistent with the qualitative observations in Figs.\ref{fig:duality}A-J.  
The quantity $\mathcal {L}_{d}=1-\mathcal {C}^2_{d}-\mathcal {D}^2_{d}$ represents the missing information, which is due to a classical lack of knowledge about the quantum system. 
The  $\mathcal {L}_{d}$  is directly related to the Tsallis $1/2$-entropy as $\mathcal {L}_{d}=\frac{1}{d-1}\left(\frac{1}{4}(S_{1/2}(\rho)+2)^2-1\right)$, 
where $S_{1/2}(\rho)=2(\text{Tr}(\rho^{1/2})-1)$ represents the Tsallis entropy.  
Note $\mathcal {L}_{d}$ is exactly the linear entropy if the $l_2$-norm coherence is used~\cite{Huber}. 
In Supplementary Sec.\ref{sec:larged} we showcase how $\mathcal {L}_{d}$ scales for a large value of $d$.

The sophisticated $d$-path interference patterns in fact contain rich information. For example, a direct quantification of the amount of coherence embedded in the $d$-path interference patterns is allowed  by measuring the visibility $\mathcal{V}_d$ (see Figs.\ref{fig:inequality}A,B and Fig.\ref{fig:FigFringeAllD}), without the need for explicitly referring to the $\rho$ of the system. 
We measured the $l_1$-norm coherence $\widetilde{\mathcal{C}}_{l1}$ for $d \in [2, 8]$, directly from the $d$-path full-wave interference fringes ($\alpha=\pi$). 
Figure \ref{fig:ArbitraryD}A shows the results that identify a linear scale-up of $\widetilde{\mathcal{C}}_{l1}$ information. 
In addition, the which-path information of $d$-outcomes is fundamentally nondeterministic and  provides a  way for quantum randomness generation. 
Based on the $d$-mode quantizations, it is possible to generate more than one bit of randomness. 
The randomness is quantified by min-entropy $H_{min} =-\log_2 p_d$, where $p_d$ is the probability of correctly guessing which-path the photons take. 
Figure \ref{fig:ArbitraryD}B shows the measured $H_{min}$ values, in the full-particle case ($\alpha=0$), generating more than one bit of randomness.

In conclusion, we have reported an experimental generalization of Bohr's duality relation in a delayed-choice multipath experiment, on a large-scale silicon-integrated quantum optical chip.  
The wave-particle transition and generalized multipath duality relation have been confirmed by the delayed choice of which-measurement performed on single photons. %
We relied on the assumptions that photons were faithfully sampled, and that the choice-maker and the observer were  independent from each other, though these assumptions could be further relaxed in the future~\cite{PhysRevLett.115.250401}. 
Our work provides a versatile  platform to study multimode quantum superposition and  coherence -- the most fundamental quantum properties and resources~\cite{coherencereview}. 
Directly probing quantum coherence from interference distributions may allow the study of quantum processes and dynamics in complex quantum physical~\cite{Leggett_2002} and biological systems~\cite{Lambert2013}.   
Going beyond the qubit-based quantum systems, highly controllable multidimensional quantum devices and systems that bases on the large-scale integrated quantum photonics platform are expected to continuously advance quantum information science and technologies~\cite{Wangreview, Elshaari2020}.     

\section*{Data Availability}
\normalsize{The data that support the findings of this study are available from the corresponding author upon reasonable request.}

\section*{Code Availability}
\normalsize{The codes that support the findings of this study are available from the corresponding author upon reasonable request.}


\section*{Acknowledgements} 
\small{We thank T. Qureshi, P. Skrzypczyk, Y. Ding and X. Yuan for useful discussions and comments. 
We acknowledge support from the National Key Research and Development (R$\&$D) Program of China (nos 2019YFA0308702, 2018YFB1107205, 2016YFA0301302), the Natural Science Foundation of China (nos 61975001, 61590933, 61904196, 61675007, 11975026, 12075159), Beijing Natural Science Foundation (Z190005), and Key R$\&$D Program of Guangdong Province (2018B030329001). S.F. acknowledges support from Shenzhen Institute for Quantum Science and Engineering, Southern University of Science and Technology (Grant No.SIQSE202005), the Key Project of Beijing Municipal Commission of Education (Grant No. KZ201810028042),  and Academy for Multidisciplinary Studies, Capital Normal University. 
M.H. acknowledges support from the Austrian Science Fund (FWF) through the START project Y789-N27. }

\section*{Authors contributions} 
\small{J.W. conceived the project. X.C., Y.D., T.P., J.M., J.B, C.Z., T. D., and H.Y. built the setup and carried out the experiment. Y.Y., B.T., and Z.L. fabricated the device. 
X.C., Y.D., S.L., T.P., J.G., S.F., M.H., and Q.H. performed the theoretical analysis. Q.H., Q.G., and J.W. managed the project. X.C., Y.D., S.L., and J.W. wrote the manuscript. 
All authors discussed the results and contributed to the manuscript. }

\section*{Competing interests} 
\small{The authors declare no competing interests.}

\newpage 
\clearpage
\pagenumbering{arabic}
\setcounter{page}{1}

\onecolumn
\section*{\centering\fontsize{15}{15}\selectfont 
Supplementary Information: \\
\vspace{0.5cm} 
A generalized multipath delayed-choice experiment on a large-scale quantum nanophotonic chip}

\begin{center}
\begin{minipage}{1\textwidth}
\begin{center}
 
\author{Xiaojiong Chen$^{1,\dagger}$, Yaohao Deng$^{1,\dagger}$, Shuheng Liu$^{1,\dagger}$, Tanumoy Pramanik$^{1,2}$, Jun Mao$^{1}$, Jueming Bao$^{1}$, Chonghao Zhai$^{1}$, \\Tianxiang Dai$^{1}$, Huihong Yuan$^{1}$,  Jiajie Guo$^{1}$, Shao-Ming Fei$^{3}$, Marcus Huber$^{4,5}$, Bo Tang$^{6}$, Yan Yang$^{6,\star}$, Zhihua Li$^{6}$, \\Qiongyi He$^{1,2,7,8,9,\star}$, Qihuang Gong$^{1,2,7,8,9,\star}$, Jianwei Wang$^{1,2,7,8,9,\star}$}
\end{center}
\end{minipage}
\end{center}

\newpage 
\clearpage

\section{Device fabrication and characterization}
\label{sec:Device}

The large-scale integrated quantum optical device for the implementation of  multipath delayed-choice experiment was designed and fabricated on the silicon nanophotonics platform, which is a versatile system for photonic quantum information processing~\cite{Wangreview}. 
The quantum device was fabricated by the standard CMOS (complementary metal-oxide-semiconductor) processes. 
A layer of photoresist was first spin on an 8-inch silicon-on-insulator (SOI) wafer with 220 nm-thick top silicon and 3 {\micro\metre}-thick buried oxide.  
The 248 nm DUV (deep ultraviolet) lithography was adopted to define the circuit patterns on photoresist as a soft mask. 
Double inductively coupled plasma (ICP) etching processes were applied to transfer the patterns from the photoresist layer to the silicon layer, forming the waveguides and circuits. 
Deep etching silicon waveguides with a etched depth of 220nm were used for the SFWM photon sources, beamsplitters and phase-shifters (see SEM image in Fig.\ref{fig:device}G). 
Shallow etching silicon waveguides with a etched depth of 70nm were used for the waveguide crossers and grating couplers (see an SEM image in Fig.\ref{fig:device}H). 
A SiO\textsubscript{2} layer of 1{\micro\metre} thickness was deposited on top of the SOI wafer by plasma-enhanced chemical vapor deposition (PECVD), working as an isolation layer between the waveguides and metal heaters to avoid potential optical losses, but maintain the high thermo-optic efficiency. 
After that, a 10 nm-thick Ti glue layer, a 20 nm-thick TiN barrier layer, a 800 nm-thick AlCu layer, and a 20 nm-thick TiN anti-reflective layer, were consequently deposited  by physical vapor deposition (PVD), and then patterned by DUV lithography and etching process to form the electrode. 
A 50 nm-thick TiN layer for thermal-optical phase-shifters was deposited and also patterned by DUV lithography and etching process. 
Finally, another 1 {\micro\metre}-thick SiO\textsubscript{2} was deposited as the top cladding layer, and followed by the bonding pad opening process.

\begin{figure}[h]
\centering
\includegraphics[width=0.95\textwidth]{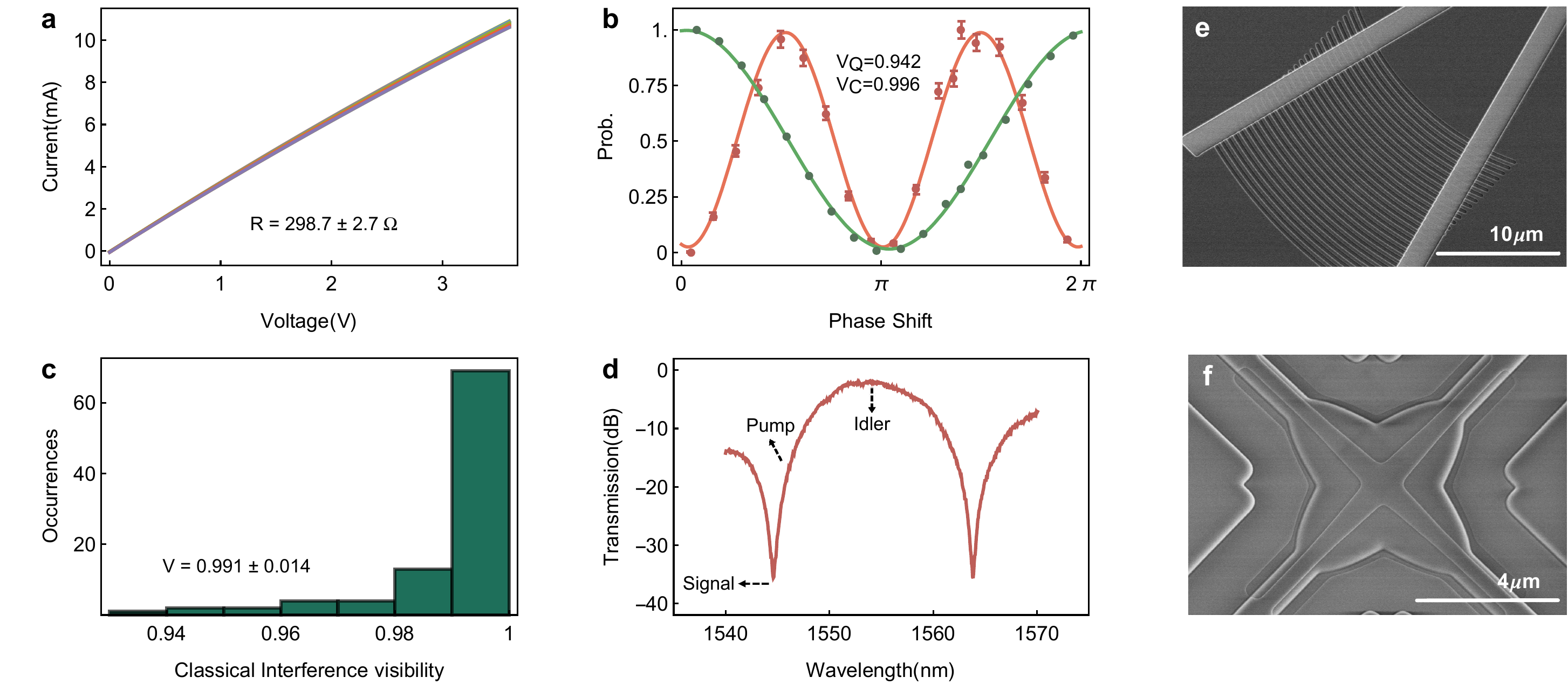}
\caption{\textbf{| Characterizations of integrated photonic components. }
\textbf{a,} Current-voltage characterization of all 95 thermo-optic phase-shifters. An averaged resistance ($R$) is measured to be $298.7\pm 2.7 \Upomega$. The fabricated heaters exhibit high uniformity of performance.  
\textbf{b,}  Single-photon classical interference fringe (green) and two-photon quantum interference fringe (red). Highly contrast visibilities are quantified. Error bars in plot are estimated from photon Poissonian statistics.   
\textbf{c,}  A histogram for all measured contrast visibilities (in total 95),  characterized from the classical interference in MZIs.  An averaged visibility of $0.991\pm 0.014$ is obtained. 
\textbf{d,}  Measured transmission spectrum for an asymmetric MZI. The spectrum is normalized to a straight waveguide with grating couplers. The wavelength for pump, signal and idlers photons are indicated by arrows. 
The asymmetric MZI allows us to separate the created signal photons and idler photons. 
\textbf{e}  and \textbf{f,} SEM  images for a fabricated grating coupler and waveguide crosser, respectively, that are both double-layer etched structures. 
}  
\label{fig:Components}
\end{figure}

\begin{figure}[t!]
\centering
\includegraphics[width=0.96\textwidth]{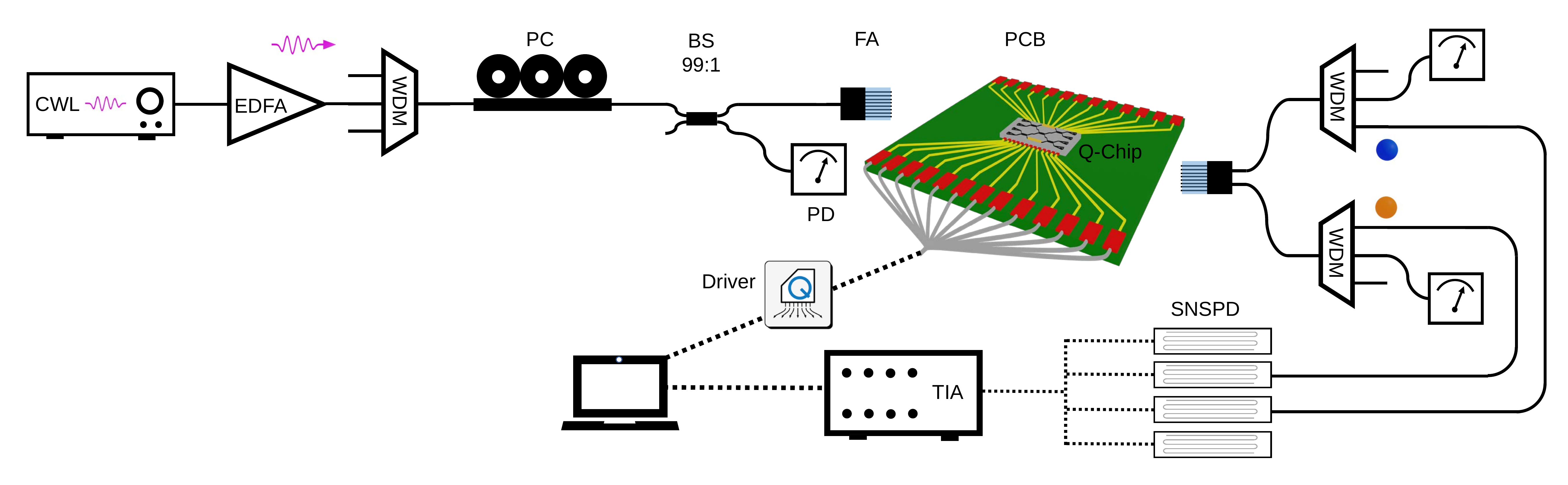}
\caption{\textbf{| Schematic of the experimental setup.} 
A continuous-wave laser (CWL, EXFO) with a wavelength of $\lambda_p=1550.11$ nm was amplified to about 40 mW by an erbium doped fiber amplifier (EDFA, Pritel). The bright laser was used as the pump source for the SFWM photon-pair generation. To remove the amplified spontaneous emission noise from the pump light, we used a dense wavelength-division multiplexing (DWDM) filter with a bandwidth of 1 nm, channel space of 1.6 nm and extinction ratio of  more than 100 dB. A fiber-based polarization controller (PC) was used to optimize the polarization before the chip, ensuing the TE (transverse electric) mode input. A  pair of path-coded entangled photons with different wavelengths, signal photon at $\lambda_{s}=1545.31$ nm and idler photon at $\lambda_{i}=1554.91$ nm, were generated via the SFWM process. The photons were manipulated and measured on-chip, by controlling the thermo-optic phase-shifters. All phase-shifters were individually accessed and driven by a multichannel voltage driver (Qontrol), with a 12-bit resolution and $\mu$s response time. The photons were then coupled off the chip into an optical fiber array (FA) for photon detection. Before that, the residual pump photon were removed from the signal and idler photons by two DWDMs, positioned after the chip. The residual pump were also monitored by photo-diodes (PDs) for the stabilization of the system.  The two single photons were ultimately detected by an array of superconducting nanowire single-photon detectors (SNSPDs, Photonspot), with a 85\% detection efficiency and 65ps time jitter. The two-fold photon coincidences were recored by a time interval analyzer (TIA, Swabian Instrument).  
The silicon quantum chip was packaged and wired bonded on a PCB (see an optical microscope image in Fig.\ref{fig:device}B). A temperature stabilization system is equipped to reduce the thermal cross-talk between heaters. 
Solid lines are optical fibers, and dotted lines are electric wires. 
 } 
\label{fig:setup}
\end{figure}

The single-mode silicon waveguides were fabricated with a cross-section of $500~\text{nm} \times 220~\text{nm}$ (full etching, see the SEM image in Fig.\ref{fig:device}G).  Multimode interferometers (MMIs) that owns large tolerance to fabrication errors, were used as 50:50 balanced beamsplitters on our chip.  The  manipulation of single photons relies on the thermal-optic tuning of phase-shifters by TiN heaters. The heaters were designed with a 100 {\micro\meter}-length and 3  {\micro\meter}-width. 
Figure \ref{fig:Components}A shows the current-voltage characterization of all 95 thermo-optic phase-shifters, showing a mean resistance of $R=298.7\pm2.7  \Omega$. Figure \ref{fig:Components}B shows the single-photon classical interference (green) and two-photon quantum interference (red), having a mean visibility of $0.996$ and $0.942$, respectively. 
The contrast visibility is defined as $({N_{max}}-{N_{min}})/({N_{max}}+{N_{min}})$), where $N$ denotes the number of photons.  
The characterization of all 2-path MZIs was performed by measuring the classical interference. 
A histogram of all measured contrast classical visibilities is shown in Fig.\ref{fig:Components}C, having a mean visibility of $0.991\pm0.014$. 
We used $d(d-1)$ 2-MZIs to realize the $d$-BS1 and $d$-BS2, which are fully reconfigurable for the choice of different dimension $d$. 
In our experiment, we used asymmetric MZIs as on-chip filters to separate the generated signal photon at $\lambda_{s}=1545.31$ nm and idler photon at $\lambda_{i}=1554.91$ nm. The asymmetric MZI filters were designed with a free spectral range (FSR) of $\lambda_{FSR}=19.2$ nm (see Fig.\ref{fig:Components}D), which are compatible to the off-chip WDM filters as shown in Fig.\ref{fig:setup}.  
The 70-nm shallowly etched grating couplers (see the SEM image in Fig.\ref{fig:device}H) in the focused configuration enabled a coupling efficiency of approx. 40\%, that allowed us to couple photons in and out of the chip.  
The insertion loss of waveguide crossers was measured to be below $-0.2$ dB and crosstalk well below $-35$ dB. See SEM images for a grating coupler and waveguide crosser in Fig.\ref{fig:Components}E and F, respectively. 
The chip was packaged on a printed circuit board (PCB), and implemented in the experimental setup, as shown in Fig~\ref{fig:setup}.

\section{Measurement of high-order interference in the $d$-path integrated interferometer}
\label{sec:highorder}

Double-path or double-slit experiments have allowed the test of Bohr's duality rule~\cite{Grangier_1986, RevModPhys.60.1067, Durr1998,C60}. In the multi-path interferometric experiment, it is of fundamental significance to rule out the existence of high-order interference terms~\cite{SORKIN}. Ruling out of the presence of such high-order interference is the basis of  testing the multipath wave-particle duality. 

We measured the magnitude of high-order interference term with respect to the second-order interference term (mutual coherence term). As an example, we characterized the fourth-order interference, by reconfiguring our device as a 4-path interferometer. The control photon was projected into the $\ket{1}$ basis, which functions as a heralding single photon in this measurement, and indicates the presence of the target photon.  The target photon was sent through the 4-path MZI for the test of  high-order interference (the four paths are labelled as $1$, $2$, $3$ and $4$, respectively). In particular, we set the $\alpha$ phase of the control photon to be $\pi$, that corresponds to the full-wave case. 
In this regard, we looked into the most obvious case of wave interference.  And, we collected the data at the prime maxima of the fringe when the $\theta$ phase was set as $\pi$ (see Fig.\ref{fig:duality}), that led to the minimal noises from photons fluctuations. 

The second-order interference term   (mutual coherence term) $I_{\uppercase\expandafter{\romannumeral2}}(ij)$ is represented as~\cite{Bohr}:

\begin{equation}
I_{\uppercase\expandafter{\romannumeral2}}(ij)=P_{\uppercase\expandafter{\romannumeral2}}(ij)-P_{\uppercase\expandafter{\romannumeral1}}(i)-P_{\uppercase\expandafter{\romannumeral1}}(j), 
\end{equation}
where $P_{\uppercase\expandafter{\romannumeral1}}(i)$ ($P_{\uppercase\expandafter{\romannumeral1}}(j)$) refers to the measured number of photons (i.e, the measured two-fold coincidence counts) when only the $i$-path ($j$-path) remains open; $P_{\uppercase\expandafter{\romannumeral2}}(ij)$ is the measured number of photons when both the $i$- and $j$-path are open; the subscripts of $\uppercase\expandafter{\romannumeral1}, \uppercase\expandafter{\romannumeral2}$ denote the number of opened modes.  

The fourth-order interference term $I_{\uppercase\expandafter{\romannumeral4}}(1234)$ can be described as~\cite{SORKIN}: 
\begin{equation}
I_{\uppercase\expandafter{\romannumeral4}}(1234)=P_{\uppercase\expandafter{\romannumeral4}}(1234)-\sum_{i < j}   P_{\uppercase\expandafter{\romannumeral2}}(ij)+2\sum_{i}   P_{\uppercase\expandafter{\romannumeral1}}(i), 
\end{equation}
where $P_{\uppercase\expandafter{\romannumeral4}}(1234)$ denotes the measured number of photons when all of the 4 paths are open. All pairs of the  $P_{\uppercase\expandafter{\romannumeral2}}(ij)$ terms, in total 6 terms, were measured by blocking all paths, but not the $i$ and $j$ paths. All of the  $P_{\uppercase\expandafter{\romannumeral1}}(i)$ terms, in total 4 terms, were measured by blocking all paths, but not the $i$ path. 
 
A normalized magnitude of $\kappa$ is adopted to represent the relative deviation from Bohr's rule. The $\kappa$ is defined as a ratio of the fourth-order interference term to the sum of all second-order interference terms~\cite{Sinha418}: 
\begin{equation}
\kappa=\frac{I_{\uppercase\expandafter{\romannumeral4}}(1234)}{\sum_{i < j}   I_{\uppercase\expandafter{\romannumeral2}}(ij)}. 
\label{Definekappa}
\end{equation}

The $\kappa$ value is expected to be zero. However, it is reasonable to be bounded by a certain accuracy, due to the presence of noises and errors in the experiment. 
Our experimental measurement returns a $\kappa=-0.0031\pm0.0047$, as shown in Fig.\ref{fig:HighOrders} in main text. This experimental result thus rules out the existence of the fourth-order interference within an accuracy of $10^{-3}$ bound, and lays the basis for the further analysis of multipath wave-particle duality. 

In our experiment, the dominant error and noise came from the Poissonian fluctuation of single photons. Note that the dark counts have been subtracted for the estimation of $\kappa$; the nonlinear compensation of  SNSPDs was not taken into account in our measurements. Another main error was introduced by the non-perfect 2-path MZIs used in our integrated nanophotonic chip. 
For example, in order to block the certain paths, we have to route away the photons by switching off the corresponding MZIs embedded in the $d$-BSs (see Fig.\ref{fig:device}E).  
That means, there could be residual photons in the blocked paths, although the probability was measured in the range of 0.01-0.001 (see the measured data in Fig.\ref{fig:Components}C). 
The cross-talk between adjacent thermal-optic phase shifters may increase the noises. 
Other contributions might come from the instability of our experimental setup, such as intensity fluctuation of the pump laser, and fiber-chip decoupling during the measurement.

\section{Multipath wave-particle quantum superposition and classical mixture}
\label{sec:DCscheme}

In this section we discuss the state evolution throughout the device, and  derive probability distributions for the quantum superposition and classical mixture  cases. 
In the next section \ref{sec:duality}, we discuss the delayed-choice of which-measurement, and through the choice of  measurement we drive the multipath duality relation. 

Photon-pair are created at the spiraled sources that are 1.4 cm-long deep etching waveguides, based on the spontaneous four-wave mixing (SFWM) nonlinear process. The dispersion of the waveguides are engineered to efficiently create photon pairs near 1550nm. Two SFWM sources are coherently pumped to  generate a bipartite state: 

\begin{equation}
 c_{0}\ket{1}_{i,0}\ket{1}_{s,0}\ket{0}_{i,1}\ket{0}_{s,1}+c_{1}\ket{0}_{i,0}\ket{0}_{s,0}\ket{1}_{i,1}\ket{1}_{s,1}, 
\end{equation}
where $\ket{1}_{i}$ ($\ket{1}_{s}$) indicates the photon number state of the idler (signal) photon being in its 0-th or 1-th spatial mode (subscripts); $\ket{0}_{i}$ ($\ket{0}_{s}$) indicates the vacuum state; $c_{0,1}$ represents the complex amplitude in each mode, having $|c_0|^2 + |c_1|^2 =1$. 
The two non-degenerate photons generated by SFWM are deterministically separated using asymmetric on-chip MZI filters and swapped by a waveguide crosser (see Fig.\ref{fig:device}).   
This results in an entangled state in the logical representation: 

\begin{equation}
c_0 \ket{0}_i\ket{0}_s+c_1 \ket{1}_i\ket{1}_s,
\end{equation}
where the coefficients $c_{0,1}$ can be chosen by arbitrary controlling the pump distribution and its phase.  Maximally path-entangled Bell states $(\ket{0}_i\ket{0}_s + e^{i\delta}\ket{1}_i\ket{1}_s) /\sqrt{2}$ can be obtained with a uniform excitation of the sources.  For simplicity, we rewrite it as: 
\begin{equation}
(\ket{0}_C\ket{0}_T + \ket{1}_C\ket{1}_T) /\sqrt{2}, 
\end{equation}
where the signal photon plays as the target while the idler photon plays as the control photon. We  locally manipulate   the state of each qubit by a SU(2) operation consisting of a MZI with an additional phase-shifter (see Fig.\ref{fig:device}). 
Then, the target photon passes through two processes coherently,  the particle-process (open $d$-path MZI, without $d$-BS2) and the wave-process (closed $d$-path MZI, with $d$-BS2), see Fig.\ref{fig:device} E. 

The choice of which-process the photon takes, either the wave or particle process, is entangled with the state of the control photon. 
Note that both processes the photon take are completely identical, except that of the $d$-BS2. 

\begin{equation}
\frac{1}{\sqrt{2}}{(\ket{0}_C\ket{P}_T+\ket{1}_C\ket{W}_T)},
\end{equation}

\vspace{-4mm}

\begin{equation}
\begin{split}
\ket{P}_T=\frac{1}{\sqrt{d}}&\sum_{m=0}^{d-1} e^{i\theta_m}\ket{m}_{T}, 
\quad
\ket{W}_T=\frac{1}{\sqrt{d}}\sum_{m=0}^{d-1} \sum_{k=0}^{d-1} h_{mk}^{(d)}e^{i\theta_k}\ket{m}_{T}, 
\label{eq:state}
\end{split}
\end{equation}

where $\ket P_{T}$ and $\ket W_{T}$ indicate the state of target photon after passing the particle process or the wave process; $\ket m$ refers to the logical state in the $k$-th path.  An arbitrary local rotation $\{\alpha, \delta\}$  is applied to the control photon, then the state evolves into: 
\begin{equation}
\frac{|0\rangle_{C}\left(\sin \frac{\alpha}{2}|P\rangle_{T}+e^{i\delta}\cos \frac{\alpha}{2}|W\rangle_{T}\right)+|1\rangle_{C}\left(\cos \frac{\alpha}{2}|P\rangle_{T}-e^{i\delta}\sin \frac{\alpha}{2}|W\rangle_{T}\right)}{\sqrt{2}}, 
\end{equation}
where $\{\alpha, \delta\}$ represent the $\{\sigma_y,\sigma_z\}$ rotations of the control photon. When we project the control photon in to the basis of $|1\rangle_C$, the state of the target photon is reduced to: 
\begin{equation}
|\psi\rangle_{T}=\cos \frac{\alpha}{2}|P\rangle_{T}-e^{i\delta}\sin\frac{\alpha}{2}|W\rangle_{T}. 
\end{equation}

Note that, up to this step, the two processes, \textit{i.e. } wave and particle processes, are classical distinguishable. 
A $d$-mode quantum erasure consisted by an array of $d$ number of 2-BSs, is ultimately adopted to erase which-process information and ensure that the wave and particle processes are in a coherent superposition as: 
\begin{equation}
|\psi\rangle_{\text {T}}=\frac{1}{\sqrt{2}}\left[\left(\cos \frac{\alpha}{2}|P\rangle_{up}-i e^{i\delta}\sin\frac{\alpha}{2}|W\rangle_{up}\right)+\left(i \cos \frac{\alpha}{2}|P\rangle_{below}-e^{i\delta}\sin \frac{\alpha}{2}|W\rangle_{below}\right)\right], 
\label{eq:AfterEraser}
\end{equation}
where $up$ ($below$) refers to the upper (below) mode of the $d$-mode quantum eraser, which are denoted at the position of $\text{D}_i (\text{D}_i')$ in Fig.\ref{fig:device}E. 
This ultimately results in the  state-process entanglement. 
 This approach of state-process entanglement  has been adopted for the implementations of controlled-unitary gate for quantum simulations~~\cite{Wang:QHL} and for the double-path delayed-choice experiments~\cite{NiceDC}.

\begin{figure*}[ht!]
\centering 
\includegraphics[width=0.745\textwidth]{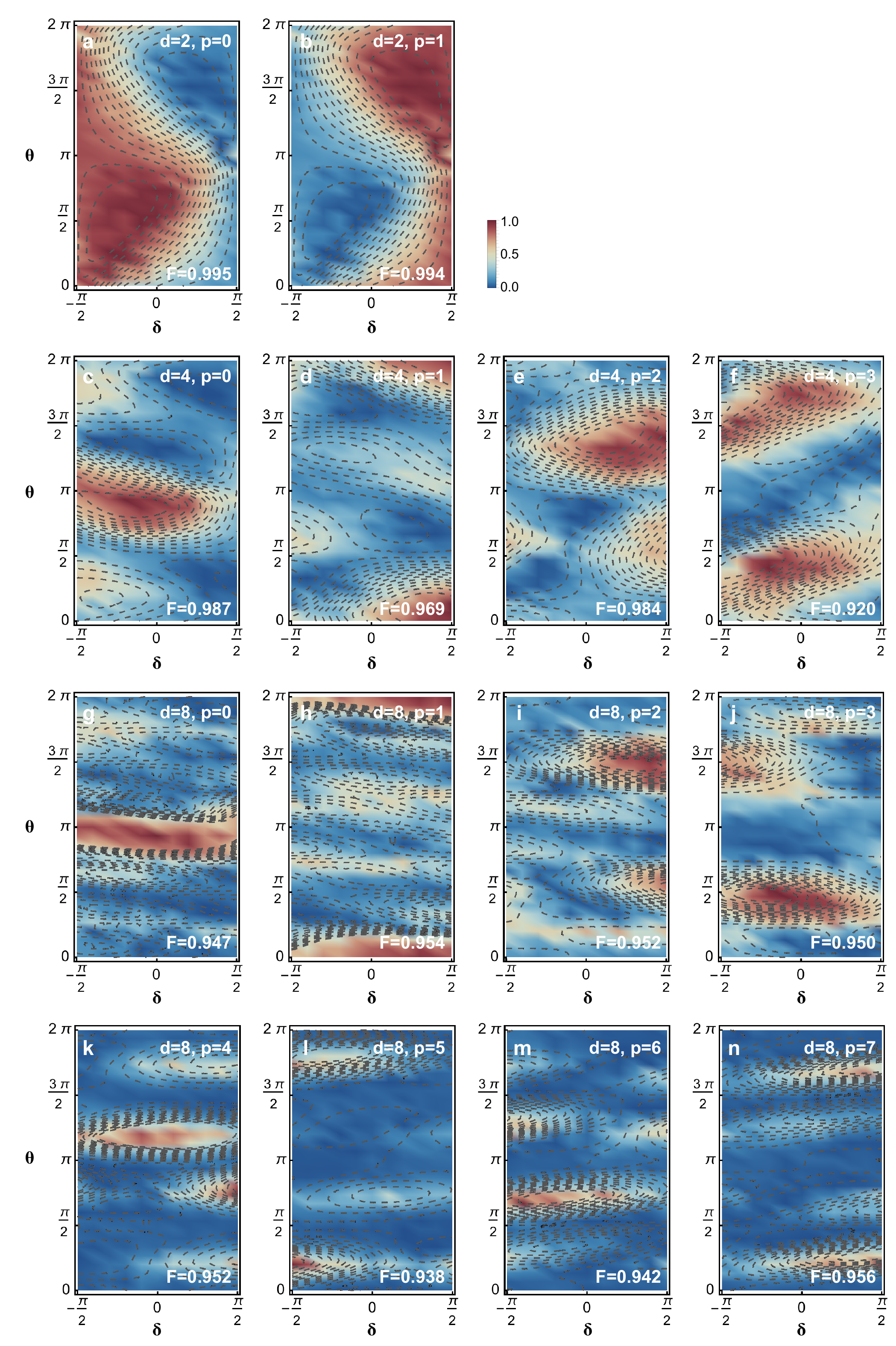}
\caption{\textbf{| Multipath quantum wave-particle transition measured at the $D_i$ mode, $ i \in [0,d-1]$. } 
The wave-particle transitions are measured in the context of genuine quantum-superposition case having $\alpha={3\pi}/{2}$, when the particle and wave nature are in maximal superposition. 
\textbf{a} and \textbf{b,} for $d=2$; \textbf{c}-\textbf{f,} for $d=4$; \textbf{g}-\textbf{n,} for $d=8$. The $p$ indicates the number of output port, and $d$ is the number of paths.  
The $\delta$-dependence of interference patterns confirms the existence of genuine wave-particle superposition. The $p=1$ data in \textbf{(a)}, \textbf{(c)} and \textbf{(g)} are shown in Fig.3, which are provided for comparison. 
Density distributions (colored) represent experimental data, while contour lines (dashed) represent theoretical results.  The ${F}$ denotes the classical fidelity, and hgh fidelities are obtained for all measurements. 
}
\label{fig:allotherports}
\end{figure*}

\begin{figure*}[ht!]
\centering 
\includegraphics[width=0.92\textwidth]{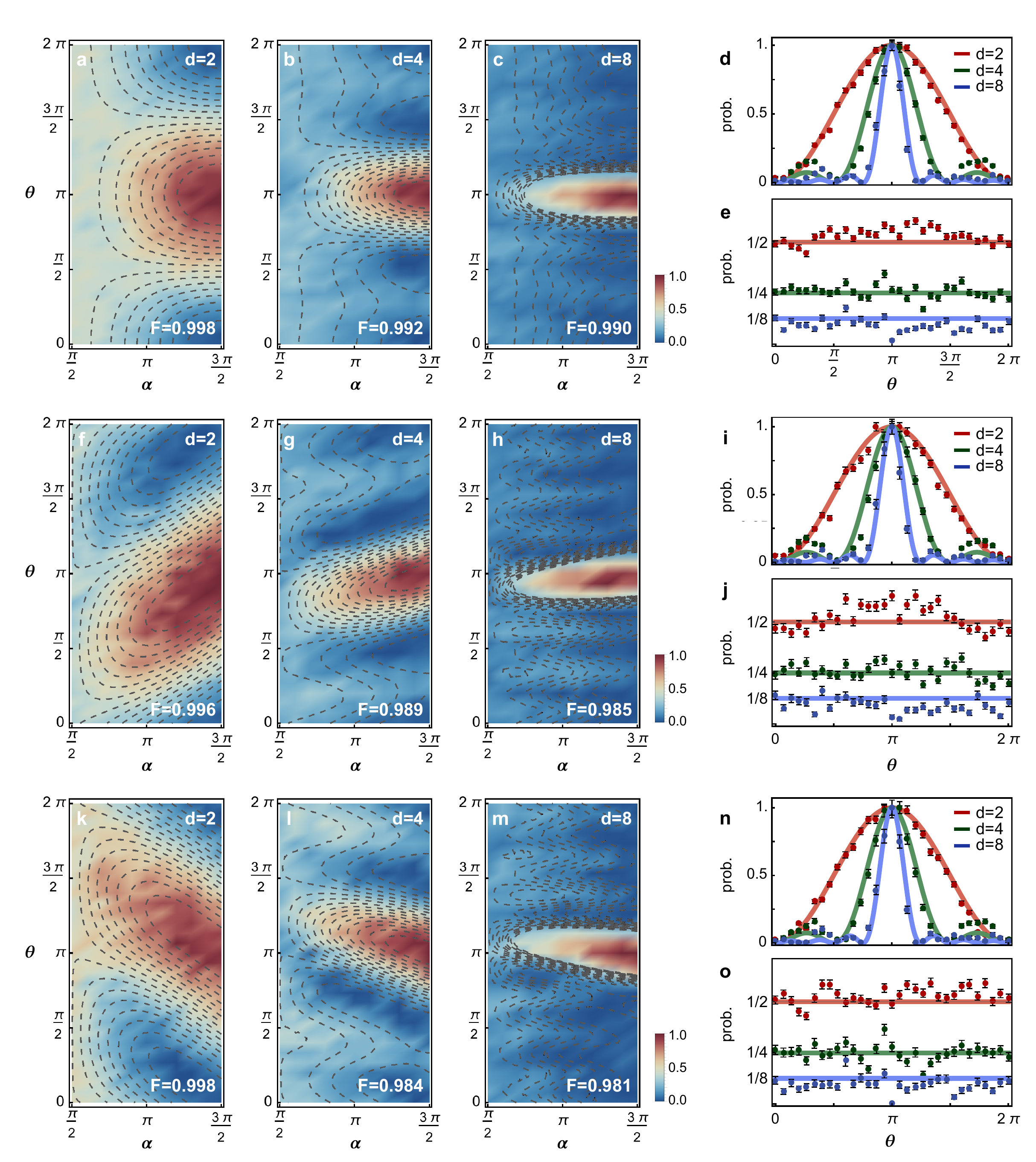}
\caption{\textbf{| Experimental results of delayed-choice multipath wave-particle transition measured in the $\sigma_x$-basis.} 
All measured data here are collected at the complementary basis $\sigma_x$, which is to demonstrate the existence of entanglement that enables the quantum delayed-choice experiment. Note the internal phase $\delta$ is chosen as zero here. Measured transitions between the particle and wave properties in three different scenarios:  \textbf{a}-\textbf{c,} classical-mixture; \textbf{f}-\textbf{h,} quantum-superposition upper mode; \textbf{k}-\textbf{m,} quantum-superposition down mode. %
They are quantified by probability distributions (normalized coincidences) for different $\alpha$ of the control and  ${\theta}_d$ of the target [${\theta}_d=k(\theta-\pi)$ was chosen], in the 2, 4 and 8-path experiments. 
Density distributions (colored) represent experimental data, while contour lines (dashed) represent theoretical results.  The ${F}$ denotes the classical fidelity $\sum_i \sqrt[]{p_iq_i}$, where $p_i$ and $q_i$ are the  measured and theoretical probabilities. High fidelities are obtained for all measurements. Results in \textbf{(a}-\textbf{c)} are consistent with classical optical multi-slit interference.  
The asymmetry of transition patterns in the quantum case \textbf{(f}-\textbf{h)} and \textbf{(k}-\textbf{m)} stem from the interference of wave and particle states. 
\textbf{d}-\textbf{e,} classical observations, \textbf{i}-\textbf{j} and \textbf{n}-\textbf{o,} quantum observations of full wave nature at $\alpha=\pi$, and full particle nature at $\alpha=0$, for $d=$2, 4 and 8. 
The interference fringe becomes sharper for $d$-path interference, while it yields multi-level quantization from $d$-outcomes of which-path information. Points represent experimental data, while lines represent theoretical values. 
All error bars ($\pm 3\sigma$) are estimated from photon Poissonian statistics. 
}
\label{fig:Xbasis}
\end{figure*}

\subsection{Quantum superposition of wave and particle}
\label{sec:quantumcase}
\noindent

When considering one of the two output modes, say $up$ mode, then the state of target photon is: 
\begin{equation}
|\psi\rangle_{\text {T}}=\frac{1}{\sqrt{N}}\left(\cos \frac{\alpha}{2}|P\rangle_{up}-i e^{i\delta}\sin\frac{\alpha}{2}|W\rangle_{up}\right),
\label{eq:UpPart}
\end{equation}
where $N$ is a normalization coefficient. Note that Eq.\ref{eq:UpPart} represents the quantum superposition of wave and particle properties. We obtain the probability of detecting the target photon in the $m$-th upper mode (after the erasure process): 
\begin{equation}
I_{m, quantum}(\hat{\theta}_d, \alpha,\delta)=\frac{1}{N}\left|\frac{1}{\sqrt{d}} \cos\frac{\alpha}{2}-ie^{i\delta}\frac{1}{\sqrt{d}}\sum_{k=0}^{d-1}h^{(d)}_{mk}e^{i\theta_k}\sin\frac{\alpha}{2}\right|^2. 
\label{eq:quantumProM}
\end{equation}
Relying on the observation of no high-order interference, the cross terms in Eq.\ref{eq:quantumProM} indicate the interference of wave and particle processes. 
For ease of representation, we choose the $d$-mode phase-shifters to be $\theta_k=k(\theta-\pi)$, with $k=0,...d-1$; however any setting of  $\{\theta_k\}^{d-1}_{k=0}$ is available. 
As an example, we consider the probability distribution of the first port, upper mode, \textit{i.e.} detected at $D_{0}$: 
\begin{equation} 
I_{0, quantum}(\hat{\theta}_d, \alpha,\delta)=\frac{1}{N}\left|\frac{1}{\sqrt{d}} \cos\frac{\alpha}{2}-i e^{i \delta}\frac{e^{i d(\theta-\pi)}-1}{d(e^{i (\theta-\pi)}-1)} \sin\frac{\alpha}{2}\right|^2 ,
\label{eq:quantumPro}
\end{equation}
with $N=1+\frac{\sin\alpha\sin\delta\sin d(\theta-\pi)}{d^{3/2} \sin(\theta-\pi)}$, which is dependent on the configuration of the control photon's $\{\alpha, \delta\}$ state. In the experiment, the probability $I_{0, quantum}$ is obtained by the normalization of two-fold coincidence, over that of all output ports.

\subsection{Classical mixture of wave and particle}
\label{sec:classicalcase}

When considering both the upper and below parts of the $d$-mode eraser and classically mix their outcomes, then the result is equivalent to the case that the which-process information is not erased and the two processes remain classical distinguishable. 
This is also equivalent to tracing out the control qubit. 
The system thus turns into the classical mixture of wave and particle: 
\begin{equation}
\rho_{T}=\cos ^{2}\frac{\alpha}{2} |P\rangle\langle P|+\sin ^{2}\frac{\alpha}{2}|W\rangle\langle W|.
\label{MixedState}
\end{equation}

The probability of detecting the target photon in the $m$-th port of eraser is given by: 
\begin{equation}
I_{m, classical}(\hat{\theta}_d, \alpha)=\frac{1}{d} \cos^{2} \frac{\alpha}{2}+\frac{1}{d}\left|\sum_{k=0}^{d-1}h_{mk}^{(d)}e^{i\theta_k}\right|^2 \sin^{2}\frac{\alpha}{2}. 
\end{equation}

In contrast to the quantum superposition case in Sec.\ref{sec:quantumcase}, the probability distribution in the classical case is $\delta$-independent as it stems from the classical mixture.  
The cross terms that represent the interference between wave and particle characters in the quantum case (Eq.\ref{eq:quantumProM}), are cancelled out -- the probability distribution $I_{m, classical}$ is a classical mixture of wave and particle properties.  
Choose the same $\theta_k=k(\theta-\pi)$, with $k=0,...d-1$, and consider the probability distribution of the first port (sum of the upper and down mode), we have: 

\begin{equation}
I_{0, classical}( \hat{\theta}_d,\alpha)=\frac{1}{d} \cos^2\frac{\alpha}{2}+\left|\frac{e^{i d(\theta-\pi)}-1}{d(e^{i (\theta-\pi)}-1)} \right|^2\sin^2\frac{\alpha}{2}.
\label{eq:classicalPro}
\end{equation}

Experimental results for both wave-particle quantum superposition and classical mixture, are shown in Fig.\ref{fig:duality} in main text (measured in the computational $\sigma_z$ basis) and Fig.\ref{fig:Xbasis} (measured in the complementary $\sigma_x$ basis). Fig. \ref{fig:allotherports} shows the measured probability distributions for other ports. 
The experimental results are in good agreement with theoretical results, reporting have level of classical fidelity.

\section{Generalized multipath duality relation in the delayed-choice experiment}
\label{sec:duality}

In this section, we represent how to construct the quantum coherence $\mathcal{C}_d$ and path-distinguishability $\mathcal{D}_d$ in theory, and how to measure them in experiment. We discuss two different scenarios,  wave-particle classical mixture and quantum superposition.    

Take the state of target photon after the $d$-BS1 as the initial state. We consider the maximal coherent state as: 
 \begin{equation}
 \ket{\Psi}_0=\frac{1}{\sqrt{d}} \sum_ {k=0}^{d-1} {\ket{k}}, 
 \end{equation}
where $\{\ket{k}\}^{d-1}_{k=0}$ is the logical basis  that defines the reference frame. Then,  $\hat{\theta}_d$ is applied on $\ket \Psi_0$. For simplicity, phases from the $d$-BS1 operator can be absorbed by the $\hat{\theta}_d$. 
To measure the genuine duality property of the target photon, not in a too naive way, the wave-particle measurement $\hat{\text{M}}_m$ is delayed chosen, which means the state of $d$-BS2 is delayed determined, by the state of \{$\alpha,\delta$\}. The choice of which-measurement $\hat{\text{M}}_m (\alpha,\delta)$ is a posterior one, projecting the state either into the full wave, full particle or their superposition basis. 

\begin{figure*}[h!]
\centering 
\includegraphics[width=0.5\textwidth]{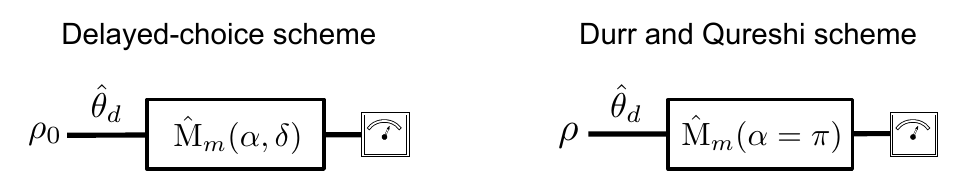}
\label{fig:Dig}
\end{figure*}

Our delayed-choice scheme is a reverse process of Durr's original scheme for measuring the $l_2$-norm coherence~\cite{Durr} and Qureshi's  scheme for measuring the $l_1$-norm coherence~\cite{Qureshicoherence}. 
In Durr's and Qureshi's scheme: given the state $\rho$, the goal is to estimate the coherence from interference patterns. 
This is allowed by performing the wave measurement $\hat{\text{M}}_m (\alpha=\pi$) on the $\rho$. 
The $\hat{\text{M}}_m (\alpha=\pi)$ is corresponding to the balanced $d$-BS operator. The probability that the photon projects to $\ket{\Psi}_{0}$ is given by: 

\begin{equation}I=
\frac{1}{d} \sum_{j=0}^{d-1} \sum_{k=0}^{d-1} \rho_{j k} e^{i\left(\theta_{j}-\theta_{k}\right)} 
=\frac{1}{d}\left(\sum_{j=0}^{d-1}\rho_{jj}+\sum_{j\neq k}\left|\rho_{jk}\right|\cos\left(\theta_j-\theta_k+\arg \rho_{jk}\right)\right)
=\frac{1}{d}\left(1+\sum_{k \neq j}\left|\rho_{j k}\right| \cos \left(\theta_{j}-\theta_{k}+\arg \rho_{j k}\right)\right). 
\label{eq:Idurr}
\end{equation}
 
Properly choosing the $\theta_{i,j}$ phases and compensating for  $\arg \rho_{j k}$,  the primary maxima $I_{max}=(1+\sum\limits_{j\neq k}|\rho_{jk}|)/d$ can be obtained. 
The generalized visibility defined as $\mathcal{V}_d=\frac{1}{d-1}\dfrac{I_{max}-I_{inc}}{I_{inc}}$, where $I_{inc}={1}/{d}$, leads to the result of~\cite{Qureshicoherence}:

\begin{equation}
\mathcal{V}_d=\mathcal{C}_d=\frac{1}{d-1}\sum_{j\neq k}\left|\rho_{jk}\right|, 
\end{equation}
where $\mathcal{C}_d$ is the normalized $l_1$-norm coherence~\cite{PlenioCoherence}. 

In our experiment, the duality relation is to be formalized in the context of  delayed-choice way.  
Our scheme is a Hermitian conjugation of Durr's scheme, which are in fact equivalent when we measure the probability distribution. 
This allows us to adopt Durr's and Qureshi's framework to quantify the coherence as well as the duality relation. 
We will discuss in detail for both classical and quantum cases as below.

\subsection{Duality relation in the quantum superposition scenario}
\label{quantumduality}

The probability  of detecting photon at the $m$-th port and upper mode, that is obtained by normalizing the measured coincidences from the detectors $\{D_{8}^{'}, D_{m}\}$, is given by: 
\begin{equation}
I_{m,quantum}=\bra m \hat{\theta}_d\rho_0\hat{\theta}_d^{\dagger}\ket m,
\label{Portm}
\end{equation}
where $\rho _0=\ket{\Psi_0}\bra{\Psi_0}$, $\hat{\theta}_d$ is the phase operator; $\ket m$ is the wave-particle basis which is defined by the state of $d$-BS2, 
\begin{equation}
\hat{\text{O}}= \cos\frac{\alpha}{2}\hat{I}-ie^{i\delta}\sin\frac{\alpha}{2}\hat{H}_d, 
\end{equation}
which is a superposition of present and absent state, controlled by \{$\alpha, \delta$\}. Note that the "absence" term ($\cos\frac{\alpha}{2}\hat{I}$) represents the measurement in the particle basis, while the "presence" term ($\sin\frac{\alpha}{2}\hat{H}_d$) represents the measurement in the wave basis. 
The state of $d$-BS2 $\hat{\text{O}}$ allows the rotation of projective measurement between the wave and particle bases, such that the maximal coherence state $\ket {\Psi_0}$ is measured in the $\ket{m}$ basis defined as: 

\begin{equation}
\ket{m}= \frac{1}{\sqrt{N_{d}}}\sum^{d-1}_{k=0}{ (\Delta_{mk} \cos \frac{\alpha}{2}+ i e^{-i\delta} h^{(d)}_{mk} \sin \frac{\alpha}{2}) \ket{k}}, 
\end{equation}
where $N_{d}=1+\sin \delta \sin \alpha/\sqrt{d}$  is the normalization coefficient; where $\Delta_x$ refers to the Kronecker function, $\Delta_{x} =1$ for $x=0$ and $\Delta_{x} =0$ for $x\neq 0$; $h^{(d)}_{mk} =\frac{1}{\sqrt{d}}(-1)^{m\odot k} $. 
At the first port upper mode, the projective basis is explicitly given by: 
\begin{equation}
\left
| m, 0\right\rangle=\frac{1}{\sqrt{N_{d}}}\left(\begin{array}{c}
	\cos \frac{\alpha}{2}+\frac{i e^{-i\delta}}{\sqrt{d}} \sin \frac{\alpha}{2} \\
	\frac{i e^{-i\delta}}{\sqrt{d}} \sin \frac{\alpha}{2} \\
	\vdots \\
	\frac{i e^{-i\delta}}{\sqrt{d}}  \sin \frac{\alpha}{2}
	\end{array}\right)
	\end{equation}

The probability can be calculated by  $I_{0,quantum}=\text{Tr}\left[\hat{\text{M}}_0 \hat{\theta}_d\rho_0\hat{\theta}_d^{\dagger} \right]$, where $\hat{\text{M}}_0=\ket{m, 0}\bra{m, 0}$. This results in the same probability distribution as Eq.\ref{eq:quantumPro}. 
Basically, our delayed-choice scheme is the Hermitian conjugation of Durr's scheme. 
Revealing either the wave or particle nature is dependent on which-measurement to perform. That means, which-measurement plays the same role as the intrinsic duality of the photon. 
Both the photon  and measurement apparatus have to be taken into account. 
In another words, the observation of duality nature of photons is reference dependent. The $d$-BS2  operator $\hat{\text{O}}$ transforms the reference-frame from the logical basis $\ket{k}$ into the wave-particle basis $\ket{m}$. 
For ease of representation, we denote $\rho$ to represent the density matrix for the entire system having the photon and delayed-choice measurement.

\subsubsection{Coherence $\mathcal{C}_d$}

The normalized $l_1$-norm coherence is given by~\cite{PlenioCoherence}: 
\begin{equation}
\mathcal{C}_d=\frac{1}{d-1}\sum_{j\neq k}\left|\rho_{jk}\right|,
\end{equation}
which is believed to be a good measure of wave nature~\cite{Bera,PRLCoherence}. We aim to obtain the coherence $\mathcal{C}_d$ from interference patterns. The  coincidence between detectors$\{D_{8}^{'}, D_{0}\} $ is normalized to obtain the probability distribution: 
 
\begin{equation}
I_{0,quantum}=\frac{1}{d}\sum_{j,k=0}^{d-1}\rho_{jk}e^{i( \theta_j-\theta_k)}          
=\frac{1}{d}\left(\sum_{j=0}^{d-1}\rho_{jj}+\sum_{j\neq k}\left|\rho_{jk}\right|\cos\left(\theta_j-\theta_k+\arg \rho_{jk}\right)\right). 
\label{eq:I1Q_rho}
\end{equation}
Note only the mutual coherence terms are present in Eq.\ref{eq:I1Q_rho}. The $I_{0,quantum}$ equals to Eq.\ref{eq:Idurr}. Choose a proper setting of $\hat{\theta}_d$ and compensate for the phases of off-diagonal elements:

\begin{equation}
\theta_0=-\arctan \frac{  \cos \frac{\alpha}{2} \cos \delta}{\frac{1}{\sqrt{d}}\sin  \frac{\alpha}{2}+\cos\frac{\alpha}{2}\sin \delta}  \hspace{10mm}  
\theta_{1}=\cdots=\theta_{d-1}=0 ,
\label{}
\end{equation}
we get the prime maxima: 

\begin{equation}
I_{max}=\frac{1}{d}(\sum_{j=0}^{d-1}\rho_{jj}+\sum_{j\neq k}\left|\rho_{jk}\right|)\\
=\frac{1}{N_{d}d}\left(\sqrt{\frac{1}{d} \sin ^{2} \frac{\alpha}{2}+\cos ^{2} \frac{\alpha}{2}+\frac{1}{\sqrt{d}} \sin \delta \sin \alpha}+\frac{d-1}{\sqrt{d}} \sin \frac{\alpha}{2}\right)^{2}.
\label{eq:imax}
\end{equation}

Note that, the prime maxima ($I_{max} $) of interference pattern consists of two parts: off-diagonal terms, representing coherence of the target photon, and diagonal terms, representing the incoherent term that have no contribution to interference $\sum_{j=0}^{d-1}\left|\rho_{jj}\right|=1$.  The incoherent term is defined as:  
\begin{equation}
I_{inc}=\frac{1}{d}\sum_{j=0}^{d-1}\rho_{jj}, 
\label{eq:inc}
\end{equation}
where 
\begin{equation}
\rho_{00}=\frac{1}{N_{d}}\left(\frac{1}{d} \sin ^{2} \frac{\alpha}{2}+\cos ^{2} \frac{\alpha}{2}+\frac{1}{\sqrt{d}} \sin \delta \sin \alpha\right), \quad \rho_{11}=\cdots=\rho_{d-1 d-1}=\frac{1}{N_{d}d} \sin ^{2} \frac{\alpha}{2}
\label{eq:rho_jj}.
\end{equation}

In our experiment, the $\rho_{ii}$ were measured as following: only the $i$ path remains open and all the other $d-1$ paths were blocked. Then the coincidences between detectors $\{D_{8}^{'}, D_{i}\}$ were measured. We measured the coincidences from all single-opening paths, and averaged them to obtain the normalized probability.  

The generalized visibility $\mathcal{V}_d$ is defined by the disparity of primary maxima $I_{max}$ and the  incoherent term $I_{inc}$~\cite{Qureshicoherence}. 
\begin{equation}
\mathcal{V}_d=\frac{1}{d-1}\frac{I_{max}-I_{inc}}{I_{inc}}. 
\label{eq:V}
\end{equation}

Substituting Eqs.\ref{eq:imax}, \ref{eq:inc} and \ref{eq:rho_jj} into Eq.\ref{eq:V}, we obtain
\begin{equation}
\mathcal{V}_d=\frac{1}{d-1}\sum_{j\neq k}\left|\rho_{jk}\right|=\mathcal{C}_d. 
\label{GeneralizedVisibility}
\end{equation}

\begin{figure*}[ht!]
\centering 
\includegraphics[width=0.98\textwidth]{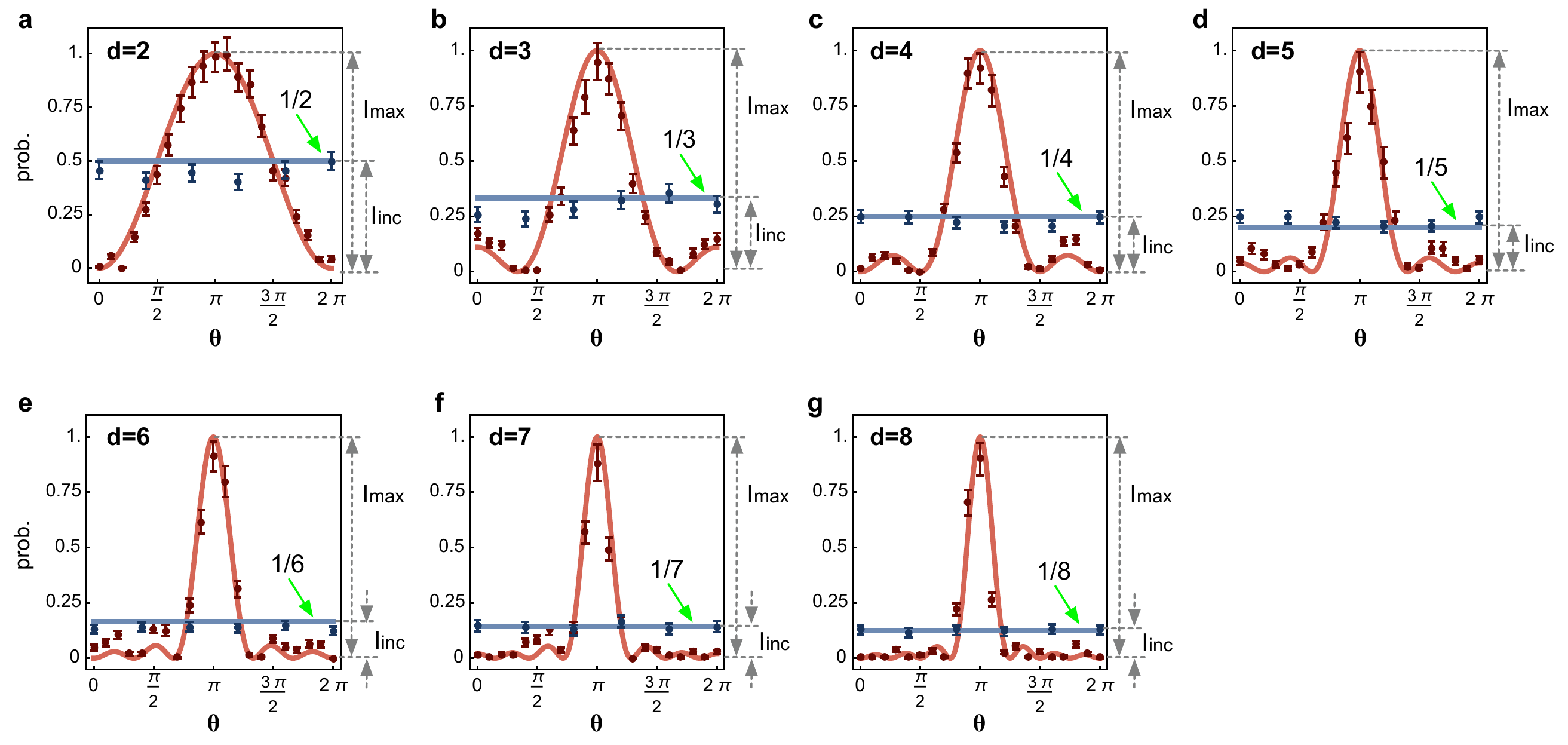}
\caption{\textbf{| Measurement of generalized visibility for $d$-path quantum interference fringes when $\alpha=\pi $}.  
We consider the $d$-path full-wave quantum interference fringes (red)  and incoherent terms (blue) for \textbf{a,}  $d=2$;  \textbf{b,}  $d=3$; \textbf{c,}  $d=4$;  \textbf{d,}  $d=5$; \textbf{e,} $d=6$; \textbf{f,} $d=7$; \textbf{g,} $d=8$. 
When $\alpha=\pi$, there is nothing to do with the interference of the two properties, so that the quantum fringes agree with classical ones. 
The generalized visibility $\mathcal {V}_{d}$  is determined by the disparity of the primary maxima $I_\text{max}$ and incoherent term $I_\text{inc}$. 
The $\mathcal {V}_{d}$ is equivalent to the quantum coherence $\mathcal {C}_{d}$, an amount of coherence information can  be directly probed from interference patterns, with no need to probing the full density matrix.  The $l_1$ norm coherence $\widetilde{\mathcal{C}}_{l1}$ for $d$-path is shown in Fig.\ref{fig:ArbitraryD}a. 
Points represent experimental data, while lines represent theoretical values. 
All error bars ($\pm \sigma$) are estimated from photon Poissonian statistics. }
\label{fig:FigFringeAllD}
\end{figure*}

Thus, we prove that the generalized visibility defined above is equivalent to the normalized $l_1$-norm coherence, \textit{i.e.} $\mathcal{V}_d=\mathcal{C}_d$.  
The result of Eq.\ref{GeneralizedVisibility} derived in our multipath delayed-choice system is the same as the one derived by Qureshi~\cite{Qureshicoherence}.  
Figure~\ref{fig:FigFringeAllD} reports the measured $l_1$-norm coherence for the $d$-path interference fringes ($d \in [2,8]$), when our device is implemented in the delayed-choice wave-particle quantum superposition.

\subsubsection{Distinguishability $\mathcal{D}_d$}

The measurement of which-path information is posteriori in our context of delayed-choice scenario. In our scheme, path-distinguishability is introduced after target photon entering the interferometer and adjusted by the state of quantum controlled $d$-BS2 -- so as the delayed choice of which-measurement, thus the which-path information we obtain is indeed posteriori~\cite{PhysRevLett.100.220402}. 

We adopted the path-distinguishability $\mathcal{D}_d$ in~\cite{Durr, Huber,QureshiD} to measure which-path information: 
\begin{equation}
\mathcal{D}_{d} = \sqrt{1-\left(\frac{1}{d-1} \sum_{i \neq j} \sqrt{\rho_{i i} \rho_{j j}}\right)^{2}}, 
\label{eq:D}
\end{equation}
which is basically a generalization of  $d=2$ distinguishability. To experimentally measure the $\mathcal{D}_{d} $, we basically measured the diagonal terms as we have discribed above.

\subsubsection{Duality relation}

Given all the definitions introduced above, we now obtain the explicit forms of $\mathcal{C}_d$ and $\mathcal{D}_d$: 
\begin{equation}
\mathcal{C}_{d}=\frac{(d-2) \sin ^{2} \frac{\alpha}{2}+2 \sin \frac{\alpha}{2} \sqrt{1+(d-1) \cos ^{2} \frac{\alpha}{2}+\sqrt{d} \sin \delta \sin \alpha}}{d+\sqrt{d} \sin \delta \sin \alpha},
\end{equation}
\begin{equation}
\mathcal{D}_{d}=\sqrt{1-\left(\frac{(d-2) \sin ^{2} \frac{\alpha}{2}+2 \sin \frac{\alpha}{2} \sqrt{1+(d-1) \cos ^{2} \frac{\alpha}{2}+\sqrt{d} \sin \delta \sin \alpha}}{d+\sqrt{d} \sin \delta \sin \alpha}\right)^2}.
\end{equation}

Bohr’s duality relation for the $d$-path interferometer can thus be formalized as: 
\begin{equation}
\mathcal{C}_d^2+\mathcal{D}_d^2\leq 1. 
\label{}
\end{equation}

Importantly, as $|\rho_{jk}|=\sqrt{|\rho_{jj}\rho_{kk}|}$, the inequality always saturates (\textit{i.e.} $\mathcal{C}_d^2+\mathcal{D}_d^2=1 $) in the context of wave-particle quantum superposition. 
In the next Section \ref{sec: classicalduality} for the case of wave-particle classical mixture, we will see that we cannot achieve the upper bound and the equality breaks up at $\alpha \neq \{0, \pi\}$, resulting in $\mathcal {C}^2_{d} + \mathcal {D}^2_{d} < 1$.

\subsection{Duality relation in the classical mixture scenario}
\label{sec: classicalduality}

Now we turn to the classical case, where the interference between wave and particle process is cancelled out. We consider the probability distribution of $m$-th port, in experiment by normalizing the coincidences between detectors $\{D_{8}^{'}, D_{m}$ and $ D_{m}^{\prime} \}$:
\begin{equation}
I_{m,classical}=\frac{\bra m \hat{\theta}_d\rho_0\hat{\theta}_d^{\dagger}\ket m+\bra {m^{\prime}} \hat{\theta}_d\rho_0\hat{\theta}_d^{\dagger}\ket {m^{\prime}}}{2},
\label{Portm}
\end{equation}
where $\ket m$ and $\ket m'$ are two wave-particle basis at the upper and bottom modes of the $m$-measurement. We consider the measurement at the first port:

\begin{equation}
\left|m, {0}\right\rangle=\left(\begin{array}{c}
	\cos \frac{\alpha}{2}+\frac{i e^{-i\delta}}{\sqrt{d}} \sin \frac{\alpha}{2} \\
	\frac{i e^{-i\delta}}{\sqrt{d}} \sin \frac{\alpha}{2} \\
	\vdots \\
	\frac{i e^{-i\delta}}{\sqrt{d}}  \sin \frac{\alpha}{2}
	\end{array}\right)
    \hspace{10 mm}
    \left|m', {0}\right\rangle=\left(\begin{array}{c}
	-i\cos \frac{\alpha}{2}-\frac{e^{-i\delta}}{\sqrt{d}} \sin \frac{\alpha}{2} \\
	-\frac{ e^{-i\delta}}{\sqrt{d}} \sin \frac{\alpha}{2} \\
	\vdots \\
	-\frac{ e^{-i\delta}}{\sqrt{d}}  \sin \frac{\alpha}{2}
	\end{array}\right),
	\label{eq:FirstMMeasurement}
    \end{equation}
    and mark $\hat{\text{M}}_0=(\ket {m,0}\bra{m,0}+\ket {m^{\prime},0}\bra{m^{\prime},0})/2$, and $I_{0,classical}=\text{Tr}\left[\hat{\text{M}}_0\hat{\theta}_d\rho_0\hat{\theta}_d^{\dagger}\right]$. The explicit form of probability is given in Eq.\ref{eq:classicalPro}.     Similar to the quantum case, the $\rho$ represents the density matrix for the whole system with the target photon and choice of measurement. 

\subsubsection{Coherence $\mathcal{C}_d$}
\noindent

The  explicit form of $I_{0,classical}$  is rewritten into: 

\begin{equation}
I_{0,classical}= \frac{1}{d}\sum_{j,k=0}^{d-1}\rho_{jk}e^{i( \theta_j-\theta_k)}          
=\frac{1}{d}\left(\sum_{j=0}^{d-1}\rho_{jj}+\sum_{j\neq k}\left|\rho_{jk}\right|\cos\left(\theta_j-\theta_k+\arg \rho_{jk}\right)\right). 
\label{}
\end{equation}

In contrast to the quantum case, here $\arg \rho_{jk}=0$. To get the prime maxima, we set $\hat{\theta}_d$ to be:

\begin{equation}
\theta_k=0  \hspace{10mm} for\hspace{2mm} k=0\hspace{1mm}...\hspace{1mm} d-1.
\end{equation}

The coincidence between detectors  $\{D_{8}^{'}, D_{0}$ and $ D_{0}^{\prime} \}$ is normalized to obtain the probability distribution, and we obtain the prime maxima: 
\begin{equation}
I_{max}=\frac{1}{d}(\sum_{j=0}^{d-1}\rho_{jj}+\sum_{j\neq k}\left|\rho_{jk}\right|)\\
=\sin^2\frac{\alpha}{2}+\frac{1}{d}\cos ^2 \frac{\alpha}{2}.
\label{eq:imaxC}
\end{equation}

The diagonal terms are given by: 
\begin{equation}
\rho_{00}=\frac{1}{d} \sin ^{2} \frac{\alpha}{2}+\cos ^{2} \frac{\alpha}{2} , \quad \rho_{11}=\cdots=\rho_{d-1 d-1}=\frac{1}{d} \sin ^{2} \frac{\alpha}{2}. 
\label{eq:rho_jj_cla}
\end{equation}

As stated in Section \ref{quantumduality}, the normalized $l_1$-norm coherence is given by the disparity of primary maxima $I_{max}$ and the  incoherent term $I_{inc}$, see Eq.\ref{eq:V}. 
Substituting Eqs.\ref{eq:imaxC} and \ref{eq:rho_jj_cla} into Eq.\ref{eq:V}, we obtain: 
\begin{equation}
\mathcal{V}_d=\frac{1}{d-1}\frac{I_{max}-I_{inc}}{I_{inc}}= \sin ^{2} \frac{\alpha}{2}=\mathcal{C}_d. 
\label{eq:GeneralizedVisibilityClassical}
\end{equation}
Note that $\mathcal{C}_d$ is independent of $d$. 

\subsubsection{Distinguishability $\mathcal{D}_d$}

Given the diagonal terms we have obtained in Eq.\ref{eq:rho_jj_cla}, and substituting Eqs.\ref{eq:rho_jj_cla}  into Eq.\ref{eq:D}, we obtain the path-distinguishability $\mathcal{D}_d$ for the classical case: 
\begin{equation}
\mathcal{D}_{d}=\sqrt{1-\left(\frac{d-2}{d} \sin ^{2} \frac{\alpha}{2}+\frac{2}{d} \sin \frac{\alpha}{2} \sqrt{1+(d-1) \cos ^{2} \frac{\alpha}{2}}\right)^2}.
\label{eq:GeneralizedDClassical}
\end{equation}

\subsubsection{Duality relation}
Given the explicit form of $\mathcal{C}_d$ and $\mathcal{D}_d$ for the classical case in Eqs.\ref{eq:GeneralizedVisibilityClassical} and \ref{eq:GeneralizedDClassical}, the duality relation for $d$-path interferometer is formalized as: 
\begin{equation}
\mathcal{C}_d^2+\mathcal{D}_d^2\leq 1. 
\label{inequality_cla}
\end{equation}

The inequality in the classical case saturates ($\mathcal{C}_d^2+\mathcal{D}_d^2=1$) only at $\alpha=\{0, \pi\}$, while it cannot achieve the upper bound and the equality breaks up at $\alpha \neq \{0, \pi\}$, resulting in $\mathcal {C}^2_{d} + \mathcal {D}^2_{d} < 1$. Experimental results for the classical duality relation are shown in Fig.\ref{fig:inequality}. Note that when $d=2$ the duality equation $ \mathcal{C}_d+\mathcal{D}_d=1$ is satisfied (see ref.~\cite{NiceDC}), but not for the generalized one with $d>2$. \\

\textbf{Lemma}\quad \textit{Pure state is the sufficient and necessary condition for duality equality $\mathcal{C}_d^2+\mathcal{D}_d^2=1$.}\\

\textbf{Proof}\quad  From the main text, it's easy to conclude that pure state is a sufficient condition.\\

If $\mathcal{C}_d^2+\mathcal{D}_d^2=1$, $|\rho_{kl}|=\sqrt{\rho_{kk}\rho_{ll}}$ is satisfied for any $(k,l)$ pair. For generality, we write state $\rho$ as $\rho=\lambda_i\rho^{(i)}$, where $\lambda\neq0$ and all $\rho^{(i)}$ are pure states. We use Einstein summation convention here. Introduce symbols $\rho_{kl}=z,\, \rho^{(i)}_{kl}=z_i,\,\rho_{kk}=x,\,\rho^{(i)}_{kk}=x_i, \,\rho_{ll}=y,\,\rho^{(i)}_{ll}=y_i$. For pure states, $|
z_i|=\sqrt{x_iy_i}$. 

On the one hand, $|z|=|\lambda_i z_i|\leq\lambda_i|z_i|$. The equation holds when all $z_i=\rho_{kl}^{(i)}$ have the same angle. On the other hand, $\sqrt{xy}=\sqrt{(\lambda_ix_i)(\lambda_jy_j)}=\sqrt{\lambda_i\lambda_jx_iy_j}$. Obviously, $x_iy_j+x_jy_i\geq2\sqrt{x_ix_jy_iy_j}$. Therefore, $\sqrt{xy}=\sqrt{\lambda_i\lambda_jx_iy_j}\geq\sqrt{\lambda_i\lambda_j\sqrt{x_ix_jy_iy_j}}=\sqrt{(\lambda_i\sqrt{x_iy_i})^2}=\lambda_i\sqrt{x_iy_i}$. The equation holds when $x_iy_j=x_jy_i$ for any $(i,j)$ pair. In other words, corresponding diagonal elements among $\rho^{(i)}$ are proportionable. 
	
According to analysis above, $\sqrt{xy}\geq\lambda_i\sqrt{x_iy_i}=\lambda_i|z_i|\geq|z|$. Consider that these two equation conditions are satisfied for any pair of $(k,l)$, we can find that $\rho^{(i)}$ are the same. Therefore, state $\rho=\lambda_i\rho^{(i)}$ must be a pure state. Then pure state is also a necessary condition.

\section{Wave-particle duality in large $d$-path interferometric experiments}
\label{sec:larged}

In this section we discuss the difference between the  wave-particle classical-mixture and  quantum-superposition cases, in a $d$-path interferometer with a large number of modes.  
In the quantum superposition case, the pure states of  target photon (given in Eq.\ref{eq:UpPart}) always saturates the duality relation of $\mathcal{C}_d^2+\mathcal{D}_d^2=1$.   
For the classical mixed states given in Eq.\ref{MixedState}, the equality breaks up at $ \alpha \neq \{0,~\pi\}$, \textit{i.e.} $\mathcal{C}_d^2+\mathcal{D}_d^2<1$. 
We define a quantity as $\mathcal {L}_{d} $ the loss of information: 
\begin{equation}
\mathcal {L}_{d} =1-\mathcal{C}^2_d-\mathcal{D}_d^2. 
\end{equation}

\begin{figure*}[ht!]
\centering 
\includegraphics[width=0.62\textwidth]{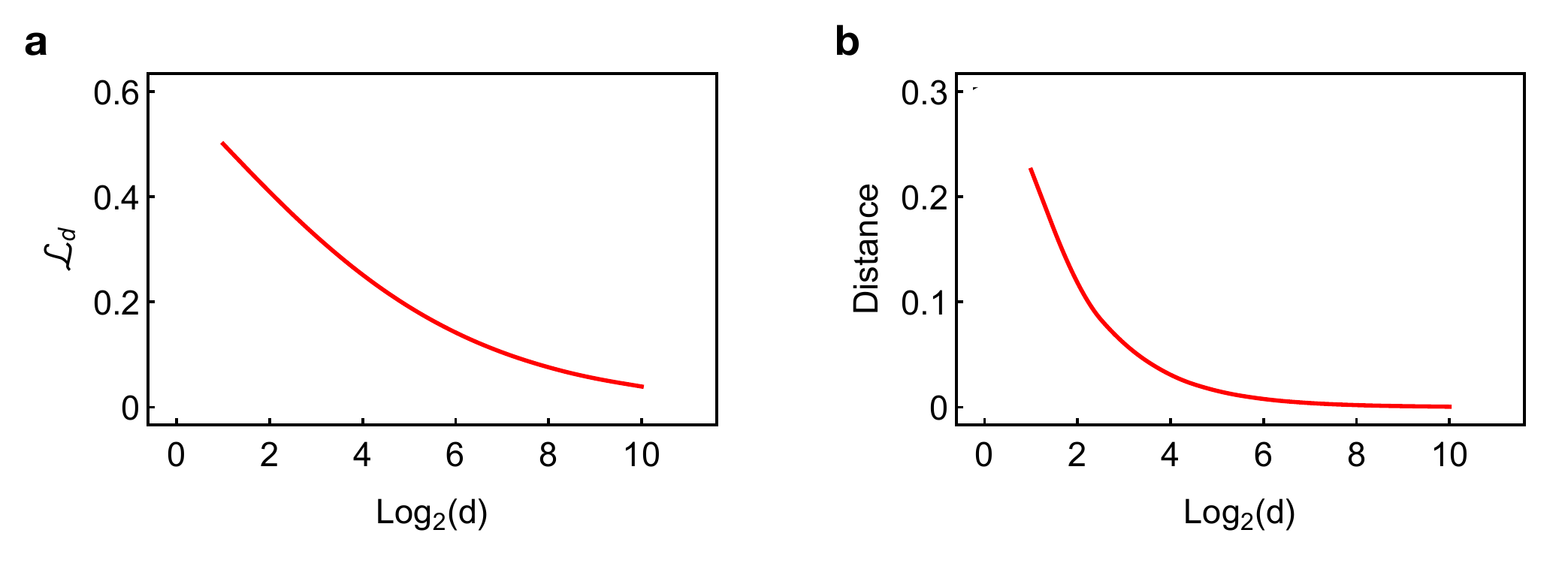}
\caption{\textbf{| Characterization of lost information and Pearson distance in a  large $d$-path interferometer. } 
\textbf{a,} The calculated $\mathcal {L}_{d}$ in the classical mixture case. $\mathcal {L}_{d}$ represents classical lack of knowledge of the quantum system.  
\textbf{b,} The calculated Pearson distance $d_{X, Y}$ between the quantum-superposition distribution and classical-mixture distribution. As the $d_{X, Y}$  approaches 0, the classical and quantum distributions tend to be same in the large $d$-path interferometer. 
}
\label{fig:LargeD}
\end{figure*}

$\mathcal {L}_{d}=0$ for  pure states. 
The $\mathcal {L}_{d}$ value is caused by tracing out the control photon, that is entangled with the target photon passing through the system. 
In the classical mixture case, the value of $\mathcal{L}_d$ is minimal at the full-particle and full-wave points at $\alpha=\{0,~\pi\}$, and maximal at certain setting of $\alpha$ (e.g, $\alpha=\pi/2$ for $d=2$), due to the quantum correlation between the control photon and target photon. 
In Fig.\ref{fig:LargeD}A, the loss of information is getting lower for the larger $d$-path system.  
The value of $\mathcal{C}_d^2+\mathcal{D}_d^2$ approaches to the unity, indicating smaller gap between the classical-mixture and quantum-superposition cases. 
Since the $l_1$-norm coherence $C_d$ for the classical case is independent of $d$ (see Eq.\ref{eq:GeneralizedVisibilityClassical}), for larger $d$, less loss of information $\mathcal{L}_d$ occurs due to the larger distinguishability of path-information $\mathcal{D}_d$  (see Eq.\ref{eq:GeneralizedDClassical}). 
This can be understood from Eq.\ref{eq:FirstMMeasurement} that in the larger $d$-MZI, the probability of detecting photons at the port $D_0$ becomes higher (when choosing the $\text{M}_0$ measurement), while the probabilities of other ports decrease, resulting in  higher distinguishability of path-information. 

We then adopt Pearson distance to quantitively describe the difference between the two distributions, that are the quantum-superposition distribution and classical-mixture distribution. 
Pearson distance is obtained by subtracting the Pearson correlation coefficient ($\rho_{X, Y}$) from $1$, 
\begin{equation}
d_{X, Y}=1-\rho_{X, Y},\quad \rho_{X, Y}=\frac{\mathcal{E}\left[\left(X-\mu_{X}\right)\left(Y-\mu_{Y}\right)\right]}{\sigma_{X} \sigma_{Y}},
\label{eq:PearsonDistance}
\end{equation}
where random variables $X$ and $Y$ represent the distributions of classical-mixture (e.g. Fig.\ref{fig:duality}A-C) and quantum-superposition (e.g. Fig.\ref{fig:duality}F-H), respectively; 
 $\mathcal{E}$, $\mu$, $\sigma$ are the expectation, mean value and standard deviation. 
As shown in Fig.\ref{fig:LargeD}B, we find that the Pearson distance tends to $0$ when the dimension $d$ increases. 
The two transition distributions behave more identical with increasing $d$. In this regard, the wave-particle quantum-superposition distribution returns to the classical-mixture distribution, in the large $d$-path interferometer.

\end{document}